\documentclass[a4paper,fleqn,usenatbib]{mnras}

\usepackage{newtxtext,newtxmath}
\usepackage[T1]{fontenc}
\usepackage{ae,aecompl}

\usepackage{epsfig}

\usepackage{graphicx}	% Including figure files
\usepackage{amsmath}	% Advanced maths commands
\usepackage{color}
\usepackage{fancybox}
\usepackage{rotating}
\usepackage{import}
\usepackage{pdfpages}
\usepackage{xcolor}
\usepackage{hyperref}
\usepackage{verbatim} 
\usepackage{CJK}

\usepackage[switch,columnwise]{lineno}
\usepackage{geometry}
\usepackage{pdflscape}
\usepackage{eqnarray}
\usepackage{soul}
\usepackage{multibib}
\newcommand{\angstrom}{{\rm \AA}}

\newcommand{\CIV}{C\,{\small IV}\,$\lambda$1549}
\newcommand{\HeII}{He\,{\small II}\,$\lambda$1640}

\newcommand{\CIII}{C\,{\small III}]\,$\lambda$1909}
\newcommand{\MgII}{Mg\,{\small II}\,$\lambda$2800}

\newcommand{\hbeta}{H{$\beta$}}

\newcommand{\objfull}{SDSS~J025214.67$-$002813.7}
\newcommand{\obj}{J0252}

\title[Periodic Quasar from Circumbinary Accretion]{Discovery of a Candidate Binary Supermassive Black Hole in a Periodic Quasar from Circumbinary Accretion Variability}
\author[Liao et al.]{
\parbox{\textwidth}{
\large
Wei-Ting Liao,$^{1,2}$\thanks{E-mail: wliao10@illinois.edu;xinliuxl@illinois.edu}
Yu-Ching Chen,$^{1,2}$
Xin Liu,$^{1,2}$
A. Miguel Holgado,$^{1,2}$
Hengxiao Guo,$^{1,2}$
Robert Gruendl,$^{1,2}$
Eric Morganson,$^{2}$
Yue Shen,$^{1,2}$\thanks{Alfred P. Sloan Research Fellow}
Tamara Davis,$^{3}$
Richard Kessler,$^{4,5}$
Paul Martini,$^{6,7}$
Richard G. McMahon,$^{8,9}$
Sahar Allam,$^{10}$
James Annis,$^{10}$
Santiago Avila,$^{11}$
Manda Banerji,$^{8,9}$
Keith Bechtol,$^{12}$
Emmanuel Bertin,$^{13, 14}$
David Brooks,$^{15}$
Elizabeth Buckley-Geer,$^{10}$
Aurelio Carnero Rosell,$^{16, 17}$
Matias Carrasco Kind$^{1, 2}$, 
Jorge Carretero$^{18}$, 
Francisco Javier Castander$^{19, 20}$, 
Carlos Cunha$^{21}$, 
Chris D'Andrea$^{22}$,
Luiz da Costa$^{17, 23}$, 
Christopher Davis$^{21}$, 
Juan De Vicente$^{16}$, 
Shantanu Desai$^{24}$, 
H. Thomas Diehl$^{10}$,
Peter Doel$^{15}$, 
Tim Eifler$^{25, 26}$,
August Evrard$^{27, 28}$, 
Brenna Flaugher$^{10}$, 
Pablo Fosalba$^{19, 20}$, 
Josh Frieman$^{10, 4}$, 
Juan Garcia-Bellido$^{29}$, 
Enrique Gaztanaga$^{19, 20}$, 
Karl Glazebrook$^{30}$, 
Daniel Gruen$^{21, 31}$, 
Julia Gschwend$^{17, 23}$, 
Gaston Gutierrez$^{10}$, 
Will Hartley$^{15, 32}$, 
Devon L. Hollowood$^{33}$, 
Klaus Honscheid$^{7, 34}$, 
Ben Hoyle$^{35, 36}$,
David James$^{37}$,
Elisabeth Krause$^{25}$, 
Kyler Kuehn$^{38}$, 
Marcos Lima$^{39, 17}$, 
Marcio Maia$^{17, 23}$,
Jennifer Marshall$^{40}$,
Felipe Menanteau$^{1, 2}$, 
Ramon Miquel$^{41, 18}$, 
Andrés Plazas Malagón$^{26}$,
Aaron Roodman$^{21, 30}$, 
Eusebio Sanchez$^{16}$, 
Vic Scarpine$^{10}$, 
Michael Schubnell$^{28}$,
Santiago Serrano$^{19, 20}$, 
Mathew Smith$^{42}$, 
R. Chris Smith$^{43}$, 
Marcelle Soares-Santos$^{44}$,
Flavia Sobreira$^{45, 17}$, 
Eric Suchyta$^{46}$, 
Molly Swanson$^{2}$, 
Gregory Tarle$^{28}$, 
Vinu Vikram$^{47}$,
Alistair Walker$^{43}$, 
the DES Collaboration
}
}

\date{Accepted XXX. Received YYY; in original form ZZZ}

\pubyear{2020}

\usepackage{eso-pic}% http://ctan.org/pkg/eso-pic

\AddToShipoutPictureBG*{%
  \AtPageUpperLeft{%
    \hspace{0.75\paperwidth}%
    \raisebox{-3.5\baselineskip}{%
      \makebox[0pt][l]{\textnormal{DES 2018-0404}}
}}}%

\AddToShipoutPictureBG*{%
  \AtPageUpperLeft{%
    \hspace{0.75\paperwidth}%
   \raisebox{-4.5\baselineskip}{%
      \makebox[0pt][l]{\textnormal{FERMILAB-PUB-20-234-AE}}
}}}%

\hypersetup{draft}

\begin{document}
%\begin{CJK*}{UTF8}{gbsn}
\label{firstpage}
\pagerange{\pageref{firstpage}--\pageref{lastpage}}
\maketitle

%\end{CJK*}

\begin{abstract}
Binary supermassive black holes (BSBHs) are expected to be a generic byproduct from hierarchical galaxy formation. The final coalescence of BSBHs is thought to be the loudest gravitational wave (GW) siren, yet no confirmed BSBH is known in the GW-dominated regime. While periodic quasars have been proposed as BSBH candidates, the physical origin of the periodicity has been largely uncertain. Here we report discovery of a periodicity ($P$=1607$\pm7$ days) at 99.95\% significance (with a global p-value of ${\sim}10^{-3}$ accounting for the look elsewhere effect) in the optical light curves of a redshift 1.53 quasar, \objfull . Combining archival Sloan Digital Sky Survey data with new, sensitive imaging from the Dark Energy Survey, the total $\sim$20-yr time baseline spans $\sim$4.6 cycles of the observed 4.4-yr (restframe 1.7-yr) periodicity. The light curves are best fit by a bursty model predicted by hydrodynamic simulations of circumbinary accretion disks. The periodicity is likely caused by accretion rate modulation by a milli-parsec BSBH emitting GWs, dynamically coupled to the circumbinary accretion disk. A bursty hydrodynamic variability model is statistically preferred over a smooth, sinusoidal model expected from relativistic Doppler boost, a kinematic effect proposed for PG1302$-$102. Furthermore, the frequency dependence of the variability amplitudes disfavors Doppler boost, lending independent support to the circumbinary accretion variability hypothesis. Given our detection rate of one BSBH candidate from circumbinary accretion variability out of 625 quasars, it suggests that future large, sensitive synoptic surveys such as the Vera C. Rubin Observatory Legacy Survey of Space and Time may be able to detect hundreds to thousands of candidate BSBHs from circumbinary accretion with direct implications for Laser Interferometer Space Antenna.
\end{abstract}

%It was targeted as a quasar candidate by the SDSS-IV/eBOSS survey based on its optical/MIR color but was not included in the SDSS DR14 quasar catalog due to its low luminosity ($M_i{=}-20.5$)
%Archival XMM-Newton X-ray observations suggest a hard X-ray luminosity of $L_{{\rm 2-10\,keV}}\approx10^{43}$ erg s$^{-1}$ (around peak) and a high level of X-ray variability, in support of an AGN as the origin for the optical variability.
% Are we sure that the X-ray variability is significant?


%%%%%%%%%%%%%%%%%%%%%%%%%%%%%%%%%%%%%%%%%%%%%%%%%%

%%%%%%%%%%%%%%%%% BODY OF PAPER %%%%%%%%%%%%%%%%%%
%%%%%%%%%%%%%%%%%%%%%%%%%%%%%%
\section{Introduction}\label{sec:intro}

LIGO has detected gravitational waves (GWs) from stellar-mass binary black hole mergers \citep{LIGO2016}, yet many GW sources are expected outside the LIGO frequency \citep{Sesana2017a,Schutz2018}. A binary supermassive black hole (BSBH) consists of two black holes, each with a mass of $\sim10^6$--$10^9$ M$_{\odot}$. BSBHs are expected to frequently form in galaxy mergers \citep{begelman80,haehnelt02,volonteri03}, given that most massive galaxies harbor SMBHs \citep{kormendy95,ff05}. Their final coalescences should produce the loudest GW sirens in the universe \citep{thorne76,haehnelt94,vecchio97,Jaffe2003}, which will be the primary source of low-frequency GW experiments \citep{Amaro-Seoane2017,Arzoumanian2018,Sesana2018}. BSBHs are important for testing general relativity in the strong field regime and for the studies of galaxy evolution and cosmology \citep{Centrella2010,DEGN,Colpi2014,Berti2015}. 

However, no confirmed case is known at sub-milliparsec scales, i.e., separations close enough to be in the GW-dominated regime. While ${\sim}150$ periodic quasars have been suggested as close BSBH candidates \citep[e.g.,][]{valtonen08,Graham2015,Charisi2016,Liutt2019,Zhu2020}, even the most promising candidates are subject to false positives due to quasar's stochastic, red noise variability, given the limited time baseline and relatively low sensitivity of existing surveys (e.g., see \citealt{Vaughan2016} for evidence against any significant periodicity in PG 1302$-$102 and \citealt{Goyal2018} in the case of OJ 287). The study of periodic quasars is important to the searches for close BSBHs in order to test theories of BSBH evolution to shed light on the expected rate of BSBH mergers as GW sources. The study is also important for understanding the physical origin of quasar periodicity, which is largely unknown.

Circumbinary accretion disks are generally expected around close BSBHs at the inferred binary separations of the candidate periodic quasars. Theory suggests that hydrodynamic variability in the circumbinary accretion disks may cause periodic light curves due to accretion rate modulation from the binary torque \citep[e.g.,][]{Farris2014,Gold2014,Shi2015,Duffell2020}. This should be useful for finding BSBHs that are close enough to be emitting gravitational waves. However, no evidence has been found for the generic ``sawtooth'' pattern (i.e., with a sharp rise and a gradual decay, in contrast to a more smooth, sinusoidal modulation expected from Doppler beaming \citep[e.g.,][]{DOrazio2015,Duffell2020}, largely limited by the relatively low sensitivity of previous surveys.

In this paper, we present a significant periodicity discovered in the optical light curves of a redshift $z$=1.53 quasar, \objfull\ (hereafter \obj\ for short). Our systematic search combines new, highly sensitive light curves from the Dark Energy Survey \citep{DarkEnergySurveyCollaboration2016} Supernova (DES-SN) fields \citep[2012--2019;][]{Bernstein2012,Goldstein2015,Kessler2015,Tie2017} with archival data from the SDSS Stripe 82 (S82) survey \citep[1998--2007;][]{ivezic07}. Unlike previous studies, which were based on few-cycle (e.g., $\sim$1.5) searches given the limited time baselines, the periodicity of \obj\ was discovered based on $\sim$5 cycles enabled by a $\sim$20-yr long baseline. The long baseline and high sensitivity are instrumental in rejecting false positives and recovering false negatives caused by stochastic quasar variability \citep[e.g.,][]{Vaughan2016,Barth2018}. Furthermore, we show that the distinct ``sawtooth'' pattern (expected from hydrodynamic circumbinary accretion disk variability models) is favored over a smoother, sinusoidal expected from Doppler beaming \citep[e.g.,][]{DOrazio2015,Duffell2020}. In addition, the frequency dependence of the variability amplitudes disfavors Doppler beaming, lending further support to the circumbinary accretion model.

%TD: In this paragraph somewhere it would be worthwhile to invest a sentence into explaining why a ?frequency-dependent variability amplitude ratio? is a prediction of the Doppler boost model. (It is not clear what that actually is, nor why the Doppler boost predicts it. If this is going to be a letter the simple things do have to be explained.

%RK: L21-29: lots of detail with big picture blurred out. You measure spectral indices in four bands, then use just 2 bands for RB model. A-ratio formula at L27 seems to be for experts. What about iz info? Suggest a wordier and more physical description here, and leave details for Method appendix.

%XL: Good point. Indeed this wasn't very well described. We do compare all possible band pairs (Extended Data Figure 3). We only quote gr bands as an example, because it is the most constraining one (ruled out at >5 sigma). The others are either less significant (only ruled out at ~2 or 3 sigma), or is consistent within uncertainties (given the larger errors on the spectral slopes).

The paper is organized as follows.  \S \ref{sec:data} describes the data and methods. \S \ref{sec:result} presents our results on the detection of a significant periodicity in \obj\ and its relevant physical properties. We discuss the implications of our results in \S \ref{sec:discuss} and conclude in \S \ref{sec:sum}.  A concordance $\Lambda$CDM cosmology with $\Omega_m = 0.3$, $\Omega_{\Lambda} = 0.7$, and $H_{0}=70$ km s$^{-1}$ Mpc$^{-1}$ is assumed throughout. We use the AB magnitude system \citep{Oke1974} unless otherwise noted.

\section{Data and Methods}\label{sec:data}

\subsection{Sample Selection} 

We start with {763} spectroscopically confirmed quasars in the {4.6} deg$^2$ overlapping region between the SDSS Stripe 82 survey and the DES-SN fields (S1 and S2). They include {758} objects in the SDSS DR7/DR14 quasar catalogs \citep{Schneider2007,Paris2018} and/or the OzDES quasar catalog \citep{Tie2017,Childress2017}, as well as {5} objects supplemented from the Million Quasars Catalog \citep{Flesch2015} (v5.5, 14 November 2018). We focus on spectroscopically confirmed quasars to ensure a clean sample in this pilot study. Only point sources are included in the analysis to avoid systematics from host galaxy contamination. We request that the DES flag SPREAD\_MODEL $<$0.005, i.e., the difference between the source point spread function (PSF) and the local PSF model is smaller than 0.5\%, or the source PSF is smaller than the local model PSF. We further require that a quasar has at least 30 $>$3$\sigma$ SDSS epochs and 50 DES epochs in at least two bands. The final parent sample consists of {625} quasars. They have a median spectroscopic redshift of {$\sim$1.8} and a median average $i$-band PSF magnitude of 21.0 mag (AB). The median epoch of observations is {80} from the SDSS and {135} from the DES.

\subsection{Light Curve Data} 

We combine archival light curves from the SDSS Stripe 82 survey with new observations from the DES-SN fields (described in detail below). The time baseline of the combined light curves extends $\sim$20 yr (1998--2007 from SDSS Stripe 82 and 2012--2019 from DES-SN). For a typical quasar at $z\sim$1, the time baseline spans $\sim10$ yr in the quasar rest-frame to encompass $\gtrsim$5 cycles for a period of $\lesssim$2 yr, which is the recommended number of cycles to minimize false periodicity \citep{Vaughan2016}. We have rejected $>5\sigma$ outliers from the running median in each band. We have binned the data within the same Julian date for a better S/N. We quote AB magnitudes throughout unless otherwise noted. 

%Describe the reduction and analysis of DES data. Pay particular attention to photometric calibration and uncertainty.

The DES is a wide-area 5000 deg$^2$ survey of the Southern Hemisphere in the $grizY$ bands \citep{Flaugher2005,TheDarkEnergySurveyCollaboration2005,DarkEnergySurveyCollaboration2016}. It uses the Dark Energy Camera \citep{Flaugher2015,Bernstein2017a} with a 2.2 degree diameter field of view mounted at the prime focus of the Victor M. Blanco 4m telescope on Cerro Tololo in Chile. The typical single-epoch 5$\sigma$ point source depths \citep{Abbott2018} are $g$=24.3, $r$=24.1, $i$=23.5, $z$=22.9, and $Y$=21.4, much deeper than other surveys of larger area (e.g., SDSS and PanSTARRS1). The data quality varies due to seeing and weather variations. The DES absolute photometric calibration has been tied to the spectrophotometric Hubble CALSPEC standard star C26202 and has been placed on the AB system \citep{Oke1983}, with an estimated single-epoch photometric statistical precision of 7.3, 6.1, 5.9, 7.3, 7.8 mmag in $grizY$ bands \citep{Abbott2018}. The DES contains a 30 deg$^2$ multi-epoch survey (DES-SN) to search for SNe Ia that has a mean cadence of $\sim$7 days in the $griz$ bands. {Two of the ten DES-SN fields (S1 and S2) are overlapped with the SDSS Stripe 82 (with an overlapping area of 4.6 deg$^2$).}  We adopt light curves generated from the Y6A1 Gold data \citep{Morganson2018}. We have also included the Science Verification data to maximize the time baseline.
%For our analysis, we use the light curves generated from the Y4A1 data (from Final Cut) supplemented with preliminary Y5 data generated using the First Cut\citep{Morganson2018}. For the Y5 data, We have double checked our zero-point calibration using the Y4A1 data as the test sample. We have verified that the results between running the First Cut and Final Cut are consistent within $\sim$0.01--0.02 mag. We have also included the Science Verification data to maximize the time baseline. 

%Describe key characteristics of the SDSS Stripe 82 data. Quote typical photometric accuracy.
The SDSS equatorial Stripe 82 region was observed from September 1998 to December 2007 with $\sim$70--90 total epochs of images in the $ugriz$ bands obtained in yearly ``seasons'' about 2--3 months long \citep{SDSSDR5,ivezic07,Frieman2008}. The typical single epoch 5$\sigma$ point source depths are 22.2, 22.2, 21.3, and 20.5 in the $griz$ bands \citep{SDSSDR5}. The photometric calibration over the survey area is accurate to roughly 0.02 mag in the $gri$ bands, and 0.03 mag in the $uz$ bands \citep{Ivezic2004}. All SDSS magnitudes have been calibrated to be nearly on the AB system \citep{SDSSDR7}.
%The random photometric errors are $<$0.01 mag for stars brighter than 20.5, 20.5, 20, and 18.5 in $griz$, respectively\citep{ivezic07}.

%RK: Is there a k correction applied to SDSS to be on DES system, or vice versa? I suspect that this effect is negligible, but a mag-limit should be given on the SDSS vs. DES comparisons.
To stitch together the light curves for each quasar, we apply the appropriate corrections to convert the SDSS photometry to be on the DES system. The corrections are to compensate for the filter coverage and system throughput differences between the two surveys. We estimate the corrections empirically by calculating two sets of synthetic magnitudes by convolving each quasar spectrum with the SDSS and DES system transmission curves (including both instrument and atmosphere). For \obj , the corrections are 
\begin{linenomath*}
\begin{equation}
\label{eq:kcorrection}
\begin{split}
g_{{\rm DES}} &=g_{{\rm SDSS}}-0.000\pm0.002 \\
r_{{\rm DES}} &=r_{{\rm SDSS}}-0.116\pm0.005 \\
i_{{\rm DES}} &=i_{{\rm SDSS}}+0.053\pm0.009 \\
z_{{\rm DES}} &=z_{{\rm SDSS}}+0.022\pm0.016, 
\end{split}
\end{equation}
\end{linenomath*}
where the errors are 1$\sigma$ uncertainties estimated from 100 bootstrap resampling of the observed quasar spectrum randomly perturbed by the error spectrum.

\begin{figure*}
\centering
\includegraphics[width=0.9\textwidth]{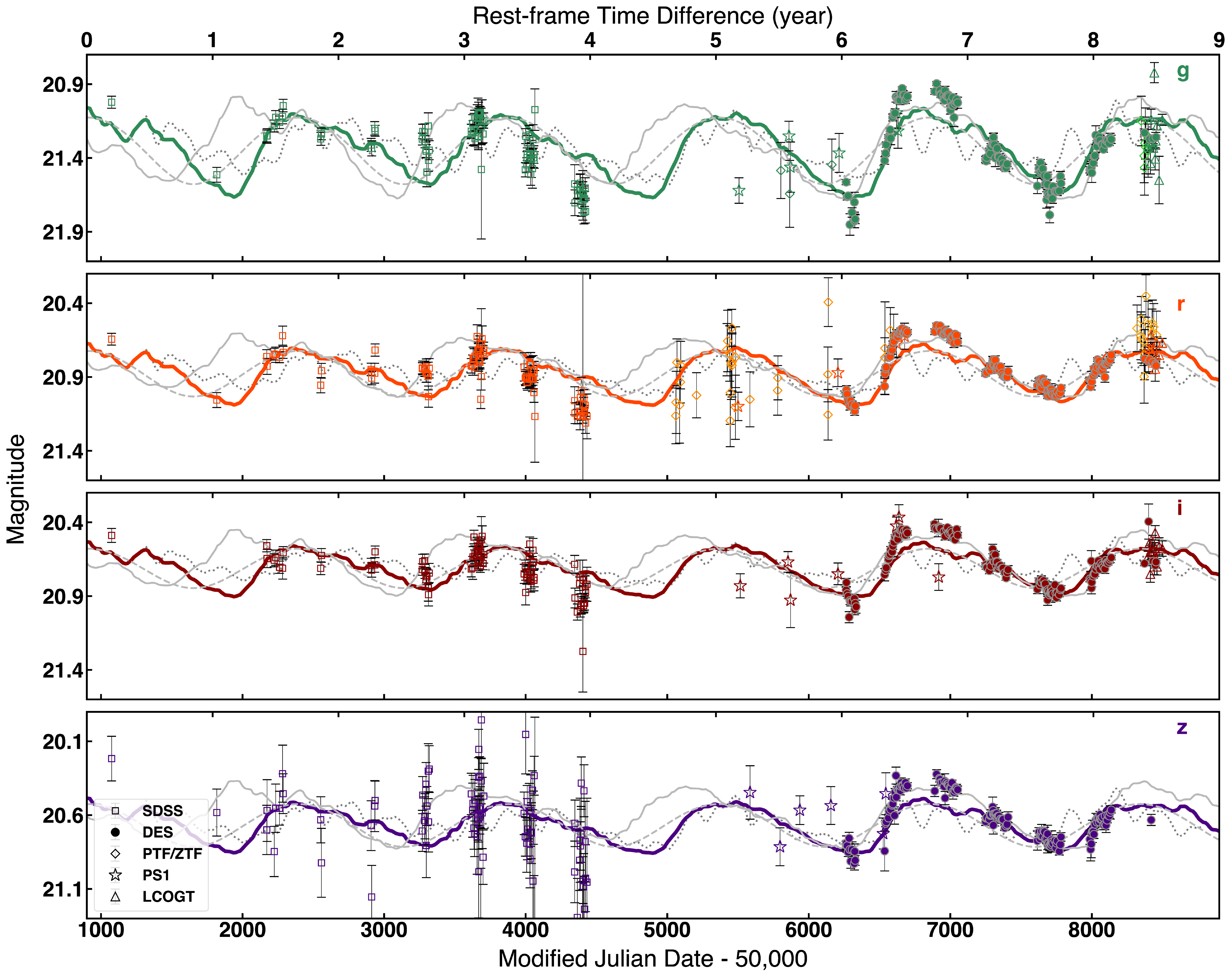}
\caption{
SDSS and DES optical light curves of \obj . All observations have been corrected to be on the DES system. Also shown are archival light curves from the PTF, CRTS, ZTF, PS1, and new observations from the LCOGT. Error bars represent 1$\sigma$ (statistical). The solid curves show the best-fit models from hydrodynamic circumbinary accretion disk simulations assuming a mass ratio $q$=0.11 \citep{Farris2014}, of which the thick solid denotes our baseline model assuming a background of random, red noise variability whereas the thin solid assumes white (flat spectrum) noise for comparison purposes only. Note that because we assume red noise in our baseline model for the background signal (from stochastic variability), the residual is not supposed to be zero, unlike the case of a white noise background. Also shown for comparison are a $q$=0.43 accretion model (dotted gray) and a sinusoidal model (dashed gray) expected from Doppler boost \citep{DOrazio2015a} both assuming red noise.
}
\label{fig:lc}
\end{figure*}

\obj\ is a spectroscopically confirmed quasar contained in the SDSS DR14 quasar catalog \citep{Paris2018}. Figure \ref{fig:lc} shows its $griz$ multi-color optical light curves. The $\sim$20-year observations combine archival SDSS data with new, higher signal to noise imaging from DES. The SDSS (DES) observations included 83, 83, 84, and 85 epochs (131, 140, 141, and 143 epochs) in the $griz$ bands with a median separation of 4 days (7 days) between epochs and yearly seasonal gaps. The variability of \obj\ is more coherent with a larger amplitude than typically observed for stochastic quasar variability \citep{Morganson2014} generally believed to result from thermal fluctuations in the accretion disks driven by an underlying stochastic process such as a turbulent magnetic field. 

Figure \ref{fig:lc} also shows archival photometry from the Palomar Transient Factory (PTF), Panoramic Survey Telescope and Rapid Response System (Pan-STARRS or PS1), and Zwicky Transient Facility (ZTF) survey, as well as our new observations from the Las Cumbres Observatory Global Telescope (LCOGT; DDT Program 2018B-004 and NOAO Program 2019A-0279; PI Liu).
%In Figure \ref{fig:lc} we also show archival photometry from the PTF survey. 
This is for the purposes of independent double checks only. We do not include them in our baseline analysis to: 1. ensure homogeneity for the analysis of the parent sample and 2. minimize possible uncertainties and/or caveats in the available data as we describe below. However, we have also verified that our results do not change qualitatively even when including them in the analysis. Further details of the archival photometry can be found in Appendix \ref{sec:archive}, with photometry data provided in the supplementary online data.

%%
%\newgeometry{margin=1cm}
%\restoregeometry
%%

%Finally, \obj\ was also covered by the Pan-STARRS1 3PI survey\citep{Chambers2016}, but the publicly available DR1 photometry\citep{Flewelling2016} was only one mean magnitude point\citep{Schlafly2012,Magnier2016}, which was averaged over some unknown (i.e., not included in the publicly available Pan-STARRS DR1, not that it was truly unknown) window of epochs, and so we could not properly include it in the analysis. We have also checked the CRTS\citep{Djorgovski2011} data archive and verified that \obj\ was too faint to be detected.

%Finally, \obj\ was also covered by the Pan-STARRS1 3PI survey\citep{Chambers2016}. Figure \ref{fig:lc} shows its mean $griz$ photometry\citep{Schlafly2012,Magnier2016} publicly available from the Pan-STARRS1 DR1\citep{Flewelling2016}. We do not include this mean magnitude point in our baseline analysis because it was averaged over some unknown (i.e., not included in the publicly available Pan-STARRS DR1, not that it was truly unknown) period of epochs, although it does follow our best-fit model. We have also checked the CRTS\citep{Djorgovski2011} data archive and verified that \obj\ was too faint to be detected. 

\subsection{Periodicity Detection}\label{subsec:periodicity} 

%We apply three steps to select candidates with a periodic variability. 
For any periodicity detection, we implement the following three selection criteria: 
\begin{enumerate}
\item At least two bands have a 3$\sigma$ detection of the same periodicity in the periodogram analysis. 
\item The detected periodicity is the {dominant} component compared to the background noise.
\item The same periodicity is also identified in the auto-correlation function (ACF). 
\end{enumerate}
%The detailed processes of each criterion are described in the following.

First, we adopt the generalized Lomb-Scargle (GLS) periodogram \citep{Zechmeister2009}, which is appropriate for detecting periodicity in unevenly sampled data using the \textit{astroML} package \citep{astroML}. Comparing the observed power to that from the simulated light curves (see details below), we identify a significant periodicity candidate if at least two bands show a 3$\sigma$ detection (i.e., the detected periodicity cannot be reproduced by >$99.7 \%$ synthetic light curves) in the same periodicity grid window. To quantify the statistical significance of any periodogram peak, we adopt an approach similar to that of \citet{Charisi2016}. Since the noise spectrum is frequency-dependent (due to the stochastic red noise quasar variability), it is more appropriate to quantify the statistical significance at a given frequency, i.e., as compared to the local background. Adopting a false-alarm probability that is flat over different frequencies instead would overestimate the true statistical significance of periodogram peaks \citep{Liutt2015}. In addition, we reject any detection where fewer than three cycles are spanned by the observations or where the periodicity is shorter than 500 days. The former criterion is imposed to minimize false periodicity due to the stochastic red noise quasar variability \citep{Vaughan2016}, whereas the latter is to mitigate artifacts caused by seasonal gaps and low cadence sampling on short timescales (i.e., an aliasing effect, e.g., \citealt{MacLeod2010}).

Second, we fit a {sine} curve to the selected candidates, and reject any of them if the residue noise dominates over the periodicity signal, i.e., if $\sigma_{\rm residue}^2 / A_{\rm sin}^2 > 1$,  where $A_{\rm sin}$ is the periodicity amplitude and $\sigma_{\rm residue}^2$ is the variance of the residue light curve after subtracting the periodic signal. 
Finally, as a complementary test, we search for periodicity by fitting the ACF with the ZDCF package \citep{Alexander1997}. For a periodicity on top of a stochastic background, ACF has a damped periodic oscillation with ACF($t$)=$\cos{(\omega t)} \exp{(-\lambda t)}$, where $t$ is the lagging time, $\omega=\frac{2 \pi}{P}$, and $\lambda$ is the decay rate of the stochastic background \citep{Graham2015}. We require that the GLS periodicity be consistent with that from the ACF test. 
Besides the above criteria, we have tested alternatives using the multi-band GLS by \cite{VanderPlas2015} and the modified GLS adopted by \cite{Zheng2016} and found that they provide no further constraint in our candidate selection, i.e., the candidates selected by the three criteria also have 3$\sigma$ detections in these alternative methods.

\subsection{Simulated Light Curves}\label{subsec:simulation} 

To quantify the statistical significance of any periodicity, first we generate 50,000 evenly sampled mocked light curves assuming a damped random walk (DRW) model with variability parameters tailored to the observed properties of each quasar. A DRW model uses a self-correcting term added to a random walk model that acts to push any deviations back toward the mean. It captures the stochastic properties of quasar variability on a timescale $\gtrsim$10 days \citep{Kelly2009,MacLeod2010,Mushotzky2011,Kozlowski2016a,Smith2018}. 
The DRW model is known as a red noise model, which has a higher spectrum power in lower frequencies. Models that failed to account for this red noise feature would likely identify false positives due to the generic power spectrum feature.  
The DRW model is governed by two parameters: $\sigma^2$ and $\tau$, which describe the flux variance and correlation timescale of the variability. 

\begin{figure*}
\centerline{
\includegraphics[width=0.9\textwidth]{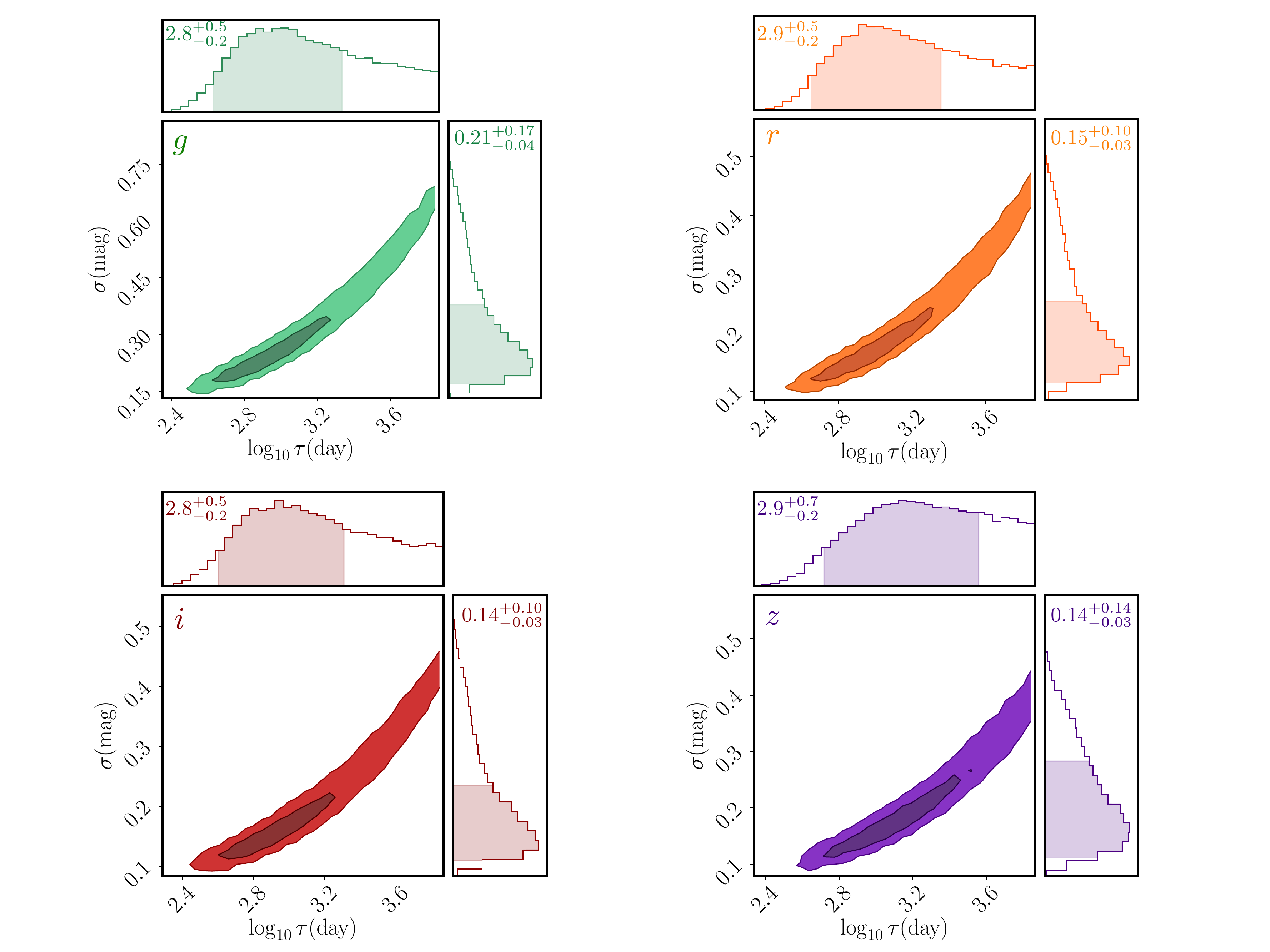}
}
\caption{DRW model parameter estimates for \obj .
The 2D contours show the 68\% and 95\% confidence levels estimated from the MCMC analysis. The histograms show the projected 1D probability density distributions for $\sigma$ and $\tau$ (observed frame). Labeled on each panel are their best-fit value and the 1-$\sigma$ (estimated from the 68\% confidence levels denoted by the shaded histograms) uncertainties. The total light curve baseline ($\sim$7300 days) is more than 10 times larger than the correlation timescale ($\sim$630 days in observed frame), so that the correlation timescale recovers the true value \citep{Kozlowski2017}.
%In the light curve simulation, the parameter pairs are randomly drawn within the two-sigma level.
}
\label{fig:drw}
\end{figure*}

To measure the DRW parameters and uncertainties for each quasar, we fit the light curve directly in the time domain by treating each data point as a state space with a Gaussian uncertainty due to both the stochastic process and measurement error, following Equations (6)--(12) of \citet{Kelly2009}. For unevenly sampled data, fitting the light curve directly in the time domain is preferred over fitting the power spectrum density for better recovering the true DRW parameters. The structure function due to the observational cadence may induce an anomalous power in the power spectrum that could potentially bias the fitting result. We apply a Bayesian model using the Markov Chain Monte Carlo (MCMC) method with the \textit{emcee} package adopting a non-informative prior \citep{Foreman-Mackey2013}. The fitting starts with 200 walkers and samples for 1500 steps. The first 750 steps are removed as a burn-in process. {To test for the convergence, we repeat the above processes but with only half of the steps (750 steps), and the resulting parameter distribution is consistent.} Figure \ref{fig:drw} shows the parameter estimation. For \obj , we have also tested a different prior with a log-normal distribution centered at 0.08 mag and 200 days for $\sigma$ and $\tau$, respectively, with a standard deviation of 1.15 \citep{Kelly2009,MacLeod2010}. The analysis is consistent across different choices of prior.   

Figure \ref{fig:drw} shows the best-fit DRW parameters for \obj . We have verified that the light curve baseline ($\sim$7300 days) is more than 10 times larger than the correlation timescale ($\sim$630 days in the observed frame), so that the correlation timescale recovers the true value \citep{Kozlowski2017}.
We then generate 50,000 mock light curves with parameters pairs ($\sigma^2$ and $\tau$) randomly drawn from the posterior distribution in the DRW parameter fitting. We down sample the mock light curves to match the cadence of the observations and add measurement errors.

We also consider a bending power law (BPL) model as an alternative to the DRW model assumption for the simulated light curves. This is motivated by results based on high-cadence \textit{Kepler} observations that suggest deviations from the DRW model at the high frequency end $f \gtrsim \frac{1}{10}$ day$^{-1}$  \citep{Mushotzky2011,Edelson2013,Edelson2014,Smith2018}. For the BPL model, we assume a $-3$ power spectrum index at the high frequency $f > \frac{1}{10}$ day$^{-1}$ and keep the DRW model at the low frequency $f < \frac{1}{10}$ day$^{-1}$. We have tested different power law indices and different high frequency breaks. Our result is insensitive to these choices. The low-frequency breaks are drawn from the timescales $\tau$ in the DRW parameter fitting. We first generate 50,000 evenly sampled mocked light curves assuming the BPL model with the \textit{pyLCSIM}\footnote{http://pabell.github.io/pylcsim/html/index.html} package. We then down sample the light curve and measure the power spectrum density using the GLS periodogram. Our result is consistent with that assuming a pure DRW model.

In addition to the DRW model, we have also considered the CAR(2,1) model \citep{Kelly2014}, i.e., a damped harmonic oscillator. CAR(2,1) is often used to describe a periodic signal \citep{Graham2015,Moreno2019}. The quality factor $Q$, defined as the ratio of the detected frequency to the corresponding frequency width, is used as a measure of the significance level. For \obj, we have $Q \ll 1$ for the CAR(2,1) model, which could suggest a low significance level for the detected period or a higher order noise.
We have tested that the significance of the periodic signal decreases when we assume the CAR(2, 1) model for the ``stochastic'' component instead of a DRW, although a $>99.74\%$ detection holds in the $g$ band. 
%However, this further strengthens our conclusion that the light curve is indeed periodic. 
Table \ref{tab:carma_BIC} shows that using \textit{carma-pack}\footnote{https://github.com/brandonckelly/carma\_pack}, CAR(1,0) has the lowest BIC value and is thus a proper noise model.  
%A DRW model is adopted for the stochastic component in our baseline noise models to differentiate it from any additional periodic component. There is no evidence for any preference for CAR(2,1) over DRW for the general quasar population, i.e., the majority of our parent quasar sample.

\begin{table*}
\centering
\begin{tabular}{l | cccccc}
\hline\hline
~~~~CAR(p, q)~~~~~~~~~~~~~ & (1,0) & (2,0) & (2,1) & (3,0) & (3,1) & (3,2)  \\
\hline
~~~~~~BIC~~~~~~~~~~~~~~ & -967 & -927 & -922 & -917 & -920 & -906 \\
\hline
\hline
\end{tabular}
\caption{ 
BIC values for CAR(p,q) model using g-band data with $p\leq 3$ and $q<p$. CAR(1,0) has the smallest BIC value, suggesting that DRW is the proper noise model for \obj.}
\label{tab:carma_BIC}
\end{table*}

%\textbf{Accounting for the look-elsewhere effect, the expected false alarm detection would be less than 2 quasars with a sample size of 625 entries. However, we want to caution that a rigorous estimation of the DRW parameter distribution over the whole quasar sample is required in order to accurately address the false detection probability, and is out of the scope of the current analysis. Although we do account for the DRW parameter distribution (Fig \ref{fig:drw}), the parameter distribution is the result of MCMC fitting on the light curve of \obj\ and is designated for \obj\ only. [Perhaps put this part in the result section.]}

%%

\subsection{eBOSS Spectrum and Analysis}\label{subsec:specdata} 

\begin{figure}
\centering
\includegraphics[width=0.5\textwidth]{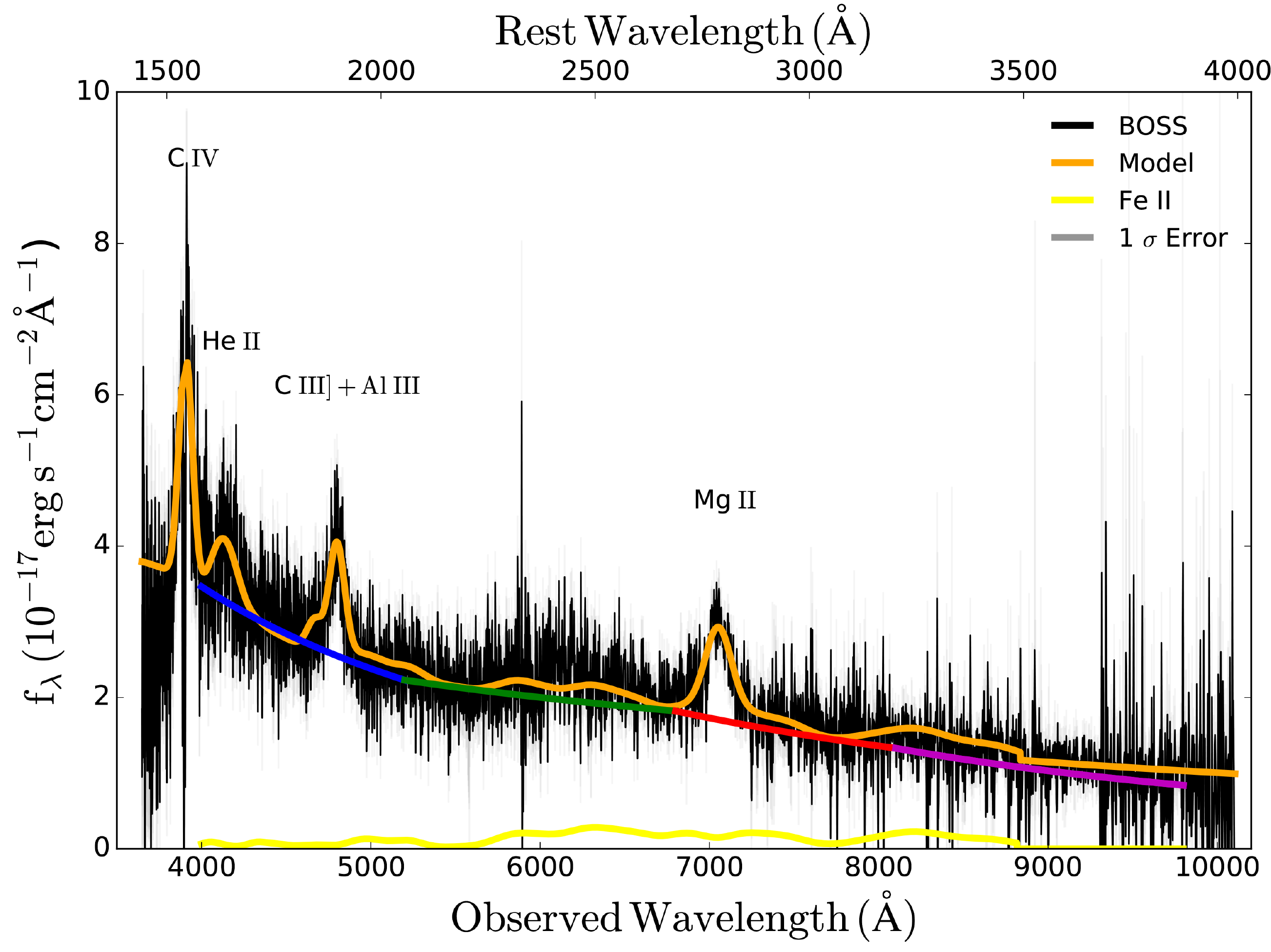}
\caption{Optical spectrum and modeling of \obj .
Shown are the data (black), the 1$\sigma$ error (gray), the best fit model (orange), the Fe {\tiny II} pseudo-continuum (yellow), and the broken power-law model for the emission-line- and Fe {\tiny II}-subtracted continuum (with the $griz$ bands plotted in blue, green, red, and magenta, respectively).
}
\label{fig:spec}
\end{figure}

\obj\ has an optical spectrum available from the SDSS DR14 data archive (Plate = 7820, Fiber ID = 470, MJD = 56984). It was taken by the BOSS spectrograph within the SDSS-IV/eBOSS survey \citep{Dawson2016}. The BOSS spectrum covers 3650--10400 \angstrom\ with a spectral resolution of $R=$1850--2200. Figure \ref{fig:spec} shows its optical (rest-frame UV) spectrum, where multiple broad emission lines are detected including \CIV , \HeII , \CIII , and \MgII .

{To estimate the viral black hole mass from the broad emission lines, we follow the procedure described in \citet{Shen2012,Shen2019} by fitting the spectral models to the observed spectra.  The spectral models contain a linear combination of power-law continuum, a pseudo continuum generated from Fe\,II emission templates, and single or multiple Gaussian components for the emission lines. Since the errors in the continuum model might change the fitting of the weak emission lines, we perform the a global fit to the mission-line free region first to construct the continuum model better. Then, we fit multiple Gaussian models to the emission lines around the \MgII\ region locally. The \MgII\ line is fitted by a combination of up to two Gaussians for the broad component and one Gaussian for the narrow component. For the FWHM of the narrow lines, we also impose an upper limit of 1200 km s$^{-1}$ . Figure \ref{fig:spec} shows our spectral models for \obj .
}

%To measure the characteristics of the broad emission lines for virial black hole mass estimates, we fit spectral models to the observed spectra following the procedures as described in details in \citep{Shen2012,Shen2019}. The model is a linear combination of a power-law continuum, a pseudo continuum constructed from Fe\,II emission templates, and single or multiple Gaussians for the emission lines. As the uncertainties in the continuum model may induce subtle effects on measurements of the weak emission lines, we first perform a global fit to the emission-line free region to better quantify the continuum components.  After subtracting the continuum we then fit multiple Gaussian models to the emission lines around the \MgII\ region locally. More specifically, we model the \MgII\ line using a combination of up to two Gaussians for the broad component and one Gaussian for the narrow component. We impose an upper limit of 1200 km s$^{-1}$ for the FWHM of the narrow lines. Figure \ref{fig:spec} shows our spectral decomposition modeling for \obj .

\section{Results}\label{sec:result}

\subsection{Discovery of A Significant Periodicity in \obj }

Using the three criteria described in \S \ref{subsec:periodicity}, we identify five significant periodic candidates out of the parent sample of {625} quasars in a {4.6} deg$^2$ field. \obj\ was the most significant detection with $>$4 cycles spanned whose light curves prefer a bursty circumbinary accretion model. We focus on \obj\ in this paper, whereas the other four candidates are presented in \citet{Chen2020}.

\begin{figure*}
\centerline{
\includegraphics[width=0.8\textwidth]{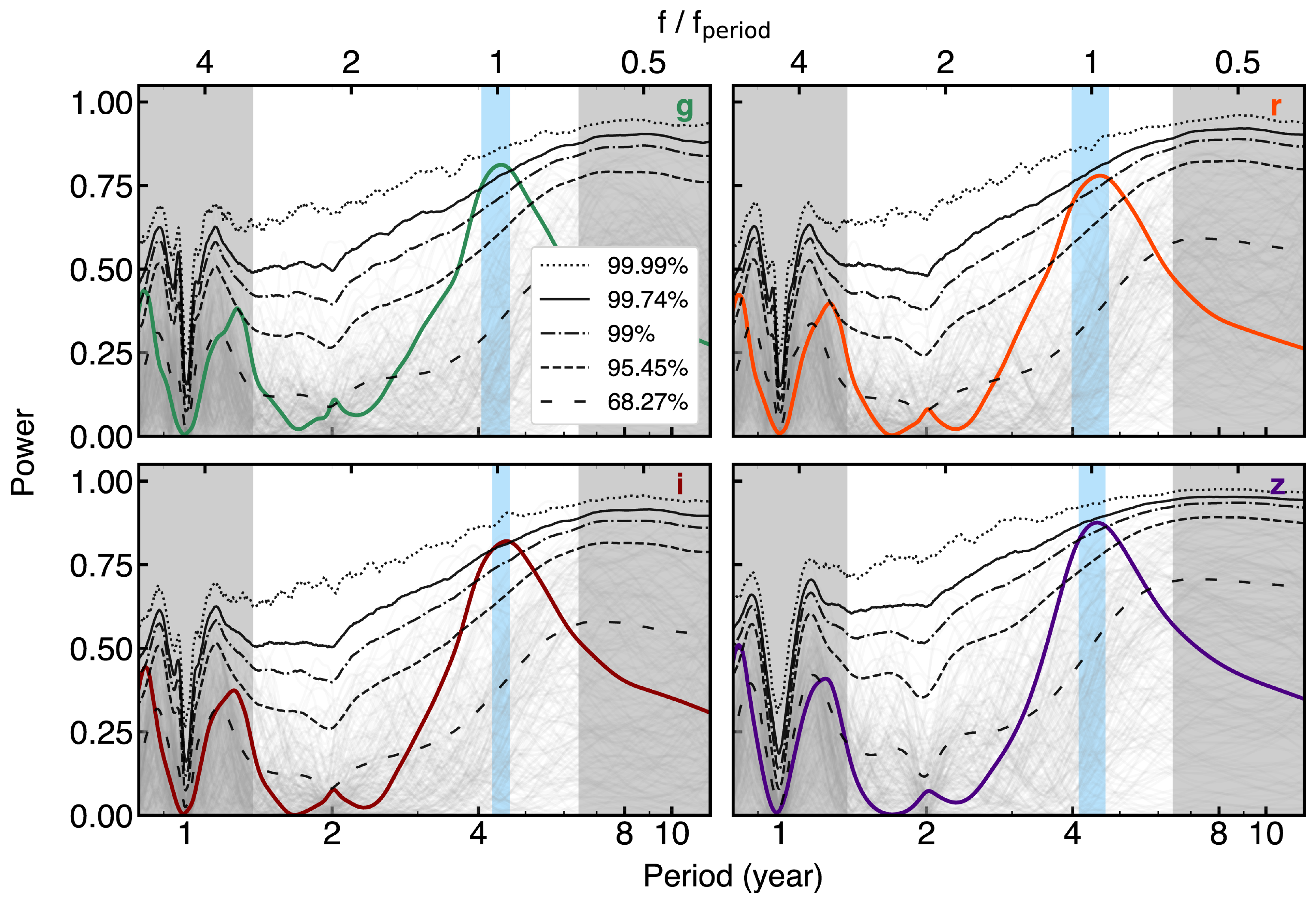}
}
\caption{Generalized Lomb-Scargle periodogram showing the periodicity detection of \obj . A periodicity (see $P_{{\rm GLS}}$ in Table \ref{tab:measurement}) is detected at 99.95\%, 99.43\%, 99.78\%, and 99.59\% single-peak significance (with global p-values of 1.2$\times10^{-3}$, 5.8$\times10^{-3}$, 2.6$\times10^{-3}$, and 8.4$\times10^{-3}$ accounting for the look elsewhere effect in the whole frequency range being searched; see \citealt{Chen2020} for details) in the $griz$ bands. The confidence levels are calculated from 50,000 tailored simulations assuming random, red noise variability. The gray curves show 200 examples drawn from the 50,000 for clarity. The cyan shaded region indicates the period uncertainty estimated using ranges above the $>$99.74\% significance for the gi bands and above the $>$99.00\% significance for the rz bands. The gray shaded regions mark the small timescales ($<$500 days) on which a periodicity may be subject to artifacts due to seasonal gaps and low cadence, and the large timescales (defined as total time baseline $<$3 cycles) where the data is more subject to false periodicity from stochastic quasar variability \citep{Vaughan2016}. 
}
\label{fig:periodogram}
\end{figure*}

\begin{figure*}
\centerline{
\includegraphics[width=0.8\textwidth]{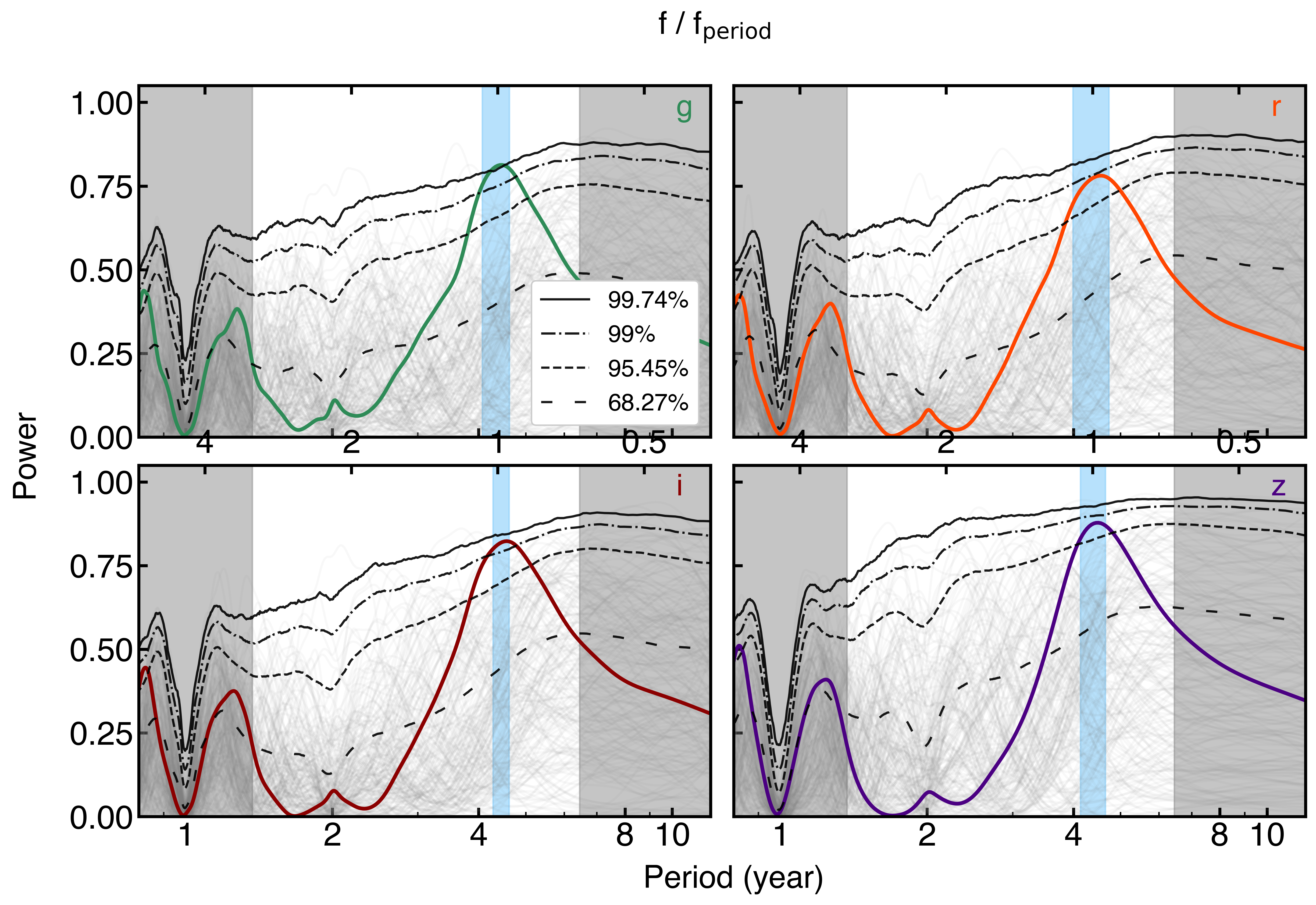}
}
\caption{Similar to Figure \ref{fig:periodogram}, but with a CAR(2,1) noise model. The cyan shaded regions are the same as those shown in Figure \ref{fig:periodogram}. The periodicity is detected at single-peak significance of $99.8$\%, $98.8$\%, $99.5$\% and $98.0$\% in the $griz$ bands.
}
\label{fig:periodogram_c21}
\end{figure*}

Figure \ref{fig:periodogram} shows the generalized Lomb-Scargle periodogram \citep{Zechmeister2009}. An observed 4.4-yr (corresponding to restframe 1.7-yr at the redshift of 1.53) periodicity is detected at 99.95\%, 99.43\%, 99.78\%, and 99.59\% single-peak significance in the $griz$ bands. The empirically estimated global p-values \citep{Barth2018} are 1.2$\times10^{-3}$, 5.8$\times10^{-3}$, 2.6$\times10^{-3}$, and 8.4$\times10^{-3}$, accounting for the look elsewhere effect \citep[e.g.,][]{Gross2010} in the whole frequency range being searched \citep[see][for details]{Chen2020}. The confidence level in each band was determined from 50,000 Monte Carlo simulations (described in \S \ref{subsec:simulation}) tailored to the observed variability flux variance and characteristic timescale assuming a damped random walk \citep[DRW;][]{Kelly2009} or a more general bending power-law (BPL) model. There is a $\sim$0.1\% probability that the periodogram peak is produced by stochastic quasar variability (i.e., assuming a correlated red noise), but the fact that we have found five candidates at $>$99.74\% single-peak significance in a parent sample of 625 (in which $\lesssim$two cases are expected from red noise; \citealt{Chen2020}) suggests that we are not just seeing stochastic quasar variability in our small sample. 
Similar to Figure \ref{fig:periodogram}, Figure \ref{fig:periodogram_c21} shows the significance level assuming a CAR(2,1) noise. The candidate periodicity is found at $99.8$\%, $98.8$\%, $99.5$\% and $98.0$\% single-peak significance in the $griz$ bands under the CAR(2,1) assumption for the stochastic noise. For context, the false alarm probability of seeing such a significant peak in the periodograms is ${\ll}10^{-20}$ assuming a pure white (i.e., flat spectrum) noise instead \citep{Zechmeister2009}. 
%\textbf{\color{red} Assuming that the DRW parameter distribution shown in Fig \ref{fig:drw} is representable to the parent sample, \citet{Chen2020} has done a rigorous estimation and address the false detection probability for candidates as a population.}
% XL: sorry but i cannot agree with your statement here. We don't make this assumption in the FAP estimate. We estimate the DRW parameters (which are different) for each quasar among the parent sample. We do the FAP estimate for each different quasar. The FAP actually varies among different quasar and also among different bands. If you have more questions, please re-read Barth and Stern (2018) on how to calculate the global FAP (here ``global'' means in the frequency space--i.e., accounting for the look elsewhere effect for any periods within the full searched range for each quasar).
\citet{Chen2020} has estimated the DRW parameter distributions for the parent quasar sample to address the false alarm probability for the candidates as a population.
%Although we do account for the DRW parameter distribution (Fig \ref{fig:drw}), the parameter distribution is the result of MCMC fitting on the light curve of \obj\ and is designated for \obj\ only.

%%

Archival observations from the PTF (in the gR bands) and from Pan-STARRS (in the griz bands) provide independent verification of our baseline observations. They also partially filled the cadence gap between the SDSS and DES observations. New observations from the LCOGT and the ZTF provide independent support and verification to our baseline DES observations. Despite having significant gaps, the combined time baseline spans ${\sim}4.6$ cycles of the periodicity, approaching the number of observed cycles recommended for minimizing false positives from stochastic quasar variability \citep[e.g.,][]{Vaughan2016}.

\subsection{Black Hole Mass Estimation}\label{subsec:bhmass} 

{The black hole mass is estimated using the single-epoch spectrum by assuming virialized motion in the broad-line region clouds \citep{Shen2013}. The broad-line region gas clouds would see the candidate BSBH as a single source. From the spectral fit to the eBOSS spectrum, the \MgII -based estimator gives a virial black hole mass of $M = 10^{8.4\pm0.1} M_{\odot}$ (1$\sigma$ statistical error), using the parameters in \citet{Vestergaard2009}. \citet{Shen2013} suggests that \MgII -based masses are more reliable than \CIV-based masses, given that \CIV\ is likely to suffer from non-virial motion like outflows and there is larger scatter between \CIV\ and \hbeta\ masses for quasars at high redshift \citep{Shen2012}. }
%We estimate the total black hole mass (because the broad-line region gas clouds would see the candidate BSBH as a single source) using the single-epoch estimator assuming virialized motion in the broad-line region clouds\citep{Shen2013}. Spectral fit to the eBOSS spectrum suggests a \MgII -based virial estimate of $M = 10^{8.4\pm0.1} M_{\odot}$ (1$\sigma$ statistical error), using the calibrations of \citep{Vestergaard2009} for \MgII . \MgII -based masses are generally considered more reliable than \CIV-based masses \citep{Shen2013}, given the larger scatter between \CIV\ and \hbeta\ masses for high-redshift quasars \citep{Shen2012}. \CIV\ is more subject to non-virial motion such as outflows. 

\subsection{Radio Loudness Upper Limit} 

\obj\ was undetected by FIRST \citep{becker95} with a 3$\sigma$ flux density upper limit of $<$0.5 mJy at 1.4 GHz. It was covered by the VLA Sky Survey \citep{VillarrealHernandez2018} (VLASS) footprint at 3 GHz to a sensitivity of 0.12 mJy RMS. It was also undetected by VLASS according to its quicklook image\footnote{http://archive-new.nrao.edu/vlass/HiPS/VLASS1.1/Quicklook/}, suggesting a 3$\sigma$ upper limit of $f^{{\rm obs}}_{{3\,{\rm GHz}}} < 0.36$ mJy. Assuming that the radio flux follows a power law $f_{\nu} \propto \nu^{\alpha}$, this translates into $f^{{\rm rest}}_{{6\,{\rm cm}}} <{0.18}$ mJy (6 cm corresponding to 5 GHz) for a spectral index $\alpha=-0.5$ \citep{Jiang2007}, or $f^{{\rm rest}}_{{6\,{\rm cm}}}<{0.20}$ mJy assuming $\alpha=-0.8$ \citep{Gibson2008}. Combining the $f_{{\rm 2500}}$ measurement from the optical spectrum, the inferred radio loudness parameter \citep{Kellermann1989} is $R\equiv f_{{6\,{\rm cm}}/f_{{\rm 2500}}} < {34}$ assuming $\alpha=-0.5$, or $R<{39}$ assuming $\alpha=-0.8$. While the VLASS upper limit cannot exclude the possibility of \obj\ being radio loud (i.e., $R>10$ according to the traditional definition based on PG quasars), it does rule out its optical emission being dominated by emission from a radio jet (i.e., $R>100$ \citep{Chiaberge2011}).

\subsection{Spectral Energy Distribution} 

\begin{figure}
\centering
\includegraphics[width=0.5\textwidth]{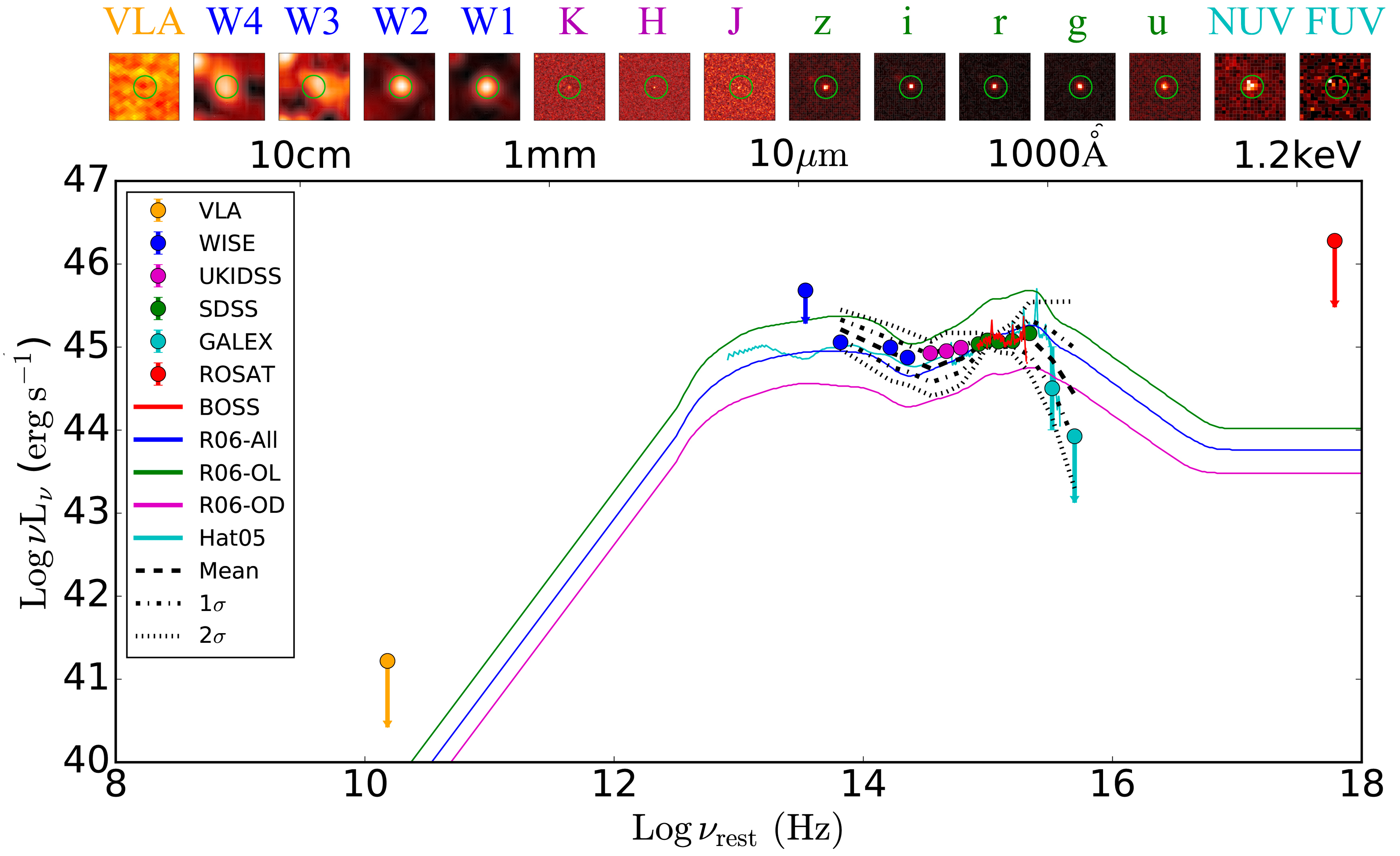}
\caption{Spectral energy distribution (SED) of \obj .
Also shown for comparison are the mean, and 1$\sigma$, and 2$\sigma$ confidence levels of the SEDs of a control quasar sample matched in redshift and luminosity with \obj , the optically selected quasar SEDs from \citep{Richards2006} (``R06-All'' for all quasars, ``R06-OL'' for optically luminous quasars, and ``R06-OD'' for optically dim quasars), and the mean SED of \citep{Hatziminaoglou2005} (Hat05). Errorbars are 1$\sigma$ whereas upper limits are 3$\sigma$. Plotted on top are the multi-wavelength postage stamps of \obj\ with a FOV of 30$''$ each. The green circles are 10$''$ in diameter indicating the position of \obj .
}
\label{fig:sed}
\end{figure}

Figure \ref{fig:sed} shows the SED of \obj . It is similar to a control sample of ordinary optically selected SDSS quasars that are matched in redshift and luminosity. The available SED observations include a radio flux density upper limit from the VLASS, MIR photometry from WISE \citep{Wright2010}, NIR photometry from UKIDSS \citep{Lawrence2007}, optical photometry from the SDSS \citep{York2000} and an optical spectrum from eBOSS, UV photometry from GALEX \citep{Martin2005} (including a detection in the NUV and an upper limit in the FUV), and an X-ray upper limit from ROSAT \citep{voges00}.

A generic prediction from circumbinary accretion disk simulations is a flux deficit in the optical/UV SED. The flux deficit may be a cutoff from a central cavity opened by the secondary black hole \citep{Milosavljevic2005} or a notch from minidisks formed around both black holes \citep{Roedig2014,Farris2015}. There is tentative evidence for an NUV deficit compared to the control sample from the existing optical spectroscopy and GALEX UV photometry, but the existing data are too uncertain to draw a firm conclusion. Future HST UV spectroscopy could confirm the potential UV deficit as a complementary test of circumbinary accretion disk models.

\section{Discussion}\label{sec:discuss}

\subsection{Physical Origins of the Periodicity}

In addition to a pure stochastic quasar variability (i.e., the null hypothesis), we consider two common, competing models for the optical light curve periodicity. The first is a smooth, sinusoidal model, which is expected from Doppler boosting. It has been proposed to explain the periodic quasar candidate PG1302$-$102 \citep{DOrazio2015a}. The highly relativistic motion of the secondary black hole drives an apparent periodicity in the light curve, assuming that the optical emission is dominated by contribution from a mini accretion disk fueling the secondary black hole. 

The second is a more bursty, quasi-periodic variability model predicted by hydrodynamic simulations of circumbinary accretion disks. We adopt the bursty hydrodynamic circumbinary accretion disk variability model of \citep{Farris2014}. The model was generated from two-dimensional (2D) hydrodynamical simulations of circumbinary disk accretion using the finite-volume code {\it DISCO} \citep{Duffell2016}. It solves the 2D viscous Navier-Stokes equations on a high-resolution moving mesh. The moving mesh shears with the fluid flow and thereby reduces the advection error in comparison to a fixed grid. Unlike previous simulations that have excised the innermost region surrounding the binary by imposing an inner boundary condition, and so potentially neglecting important dynamics occurring inside the excised region, the model was the first 2D study to include the inner cavity using shock-capturing Godunov-type methods. The simulations last longer than a viscous time such that the solutions represent a quasi-steady accretion state. 

More specifically we consider two models, mass ratio $q$=0.11 and $q$=0.43. These two values are chosen because they represent two characteristic regimes in the light-curve behaviors (Figure 9 of \citep{Farris2014}). In the simulations, significant periodicity in the accretion rates emerges only for $q\gtrsim$0.1, where the binary torques are large enough to excite eccentricity in the inner cavity and create an overdense lump. The passing BHs interact with the overdense lump, producing periodicity in the accretion rate. There is a strong peak in the periodograms corresponding to the orbital frequency of the lump, which is also the binary frequency for $q$=0.11 but is $\sim$1/5 of the binary frequency for $q$=0.43. The quality of the existing light curves does not justify model comparison over an even finer parameter grid in mass ratio. 

One caveat is that the 2D models only predict the accretion rate and miss 3D effects and radiative transfer processes. While more realistic simulations are still needed to capture the complex physics in the binary system in order to make reliable predictions, the dominant characteristic timescale, the orbital period, and harmonics that might arise, should emerge in the light curve. The gas has to be accelerated by the binary potential, and the emission of the gas has to reflect, at some level, this behavior. Whether or not we can get an accurate estimate of the mass ratio is indeed uncertain, but circumbinary accretion variability is still preferred over relativistic Doppler boosting both for the more bursty light curve characteristic and the frequency dependent variability amplitudes as discussed further below.

%To draw a parallel from another topic for example, EHT modelers perform GRMHD simulations that have magnetic fields, and post-processed radiative transfer. However, these models still lack detailed plasma physics processes and self-consistent radiative transfer that feeds back on the plasma dynamics. These models can still be used to generate images to compare with EHT data and the resulting image should ultimately have the dominant features: a black-hole shadow, with relativistic boosting from the gas motion relative to the line-of-sight. 

%For our case of SMBH binary candidates, the dominant feature we seek are periodicity that may be consistent with orbital motion, which should be the dominant feature in circumbinary accretion disks. 

\subsection{Light Curve Model Fitting and Model Comparison} 

We have shown that a periodic model is preferred over a correlated red noise (i.e., modeled with a DRW model) based on the periodogram analysis using tailored simulations (Figure \ref{fig:periodogram}). As an independent analysis, here we also fit the light curve with a covariance matrix that includes a correlated red noise between measurements. It allows us to test if the data favors an additional periodic signal on top of a background of pure random, red noise variability (i.e., from stochastic quasar variability), as well as to perform a comparison between a smooth, sinusoidal model and the more bursty accretion models. 

The covariance matrix between measurements is given by
\begin{linenomath*}
\begin{equation}
C_{i j} = 
{\sigma_i}^2 \delta_{i j} + 
\sigma^2 \exp{\bigg[\frac{- |t_i - t_j |}{\tau}\bigg]} \, ,
\end{equation}
\end{linenomath*}
%\begin{linenomath*}
%\begin{equation}
%    C_{i j}=
%    \begin{cases}
%      {\sigma_i}^2, & \text{if}\ i=j \\
%      \sigma^2 \exp{\bigg[\frac{- |t_i %- t_j |}{\tau}\bigg]}, & %\text{otherwise }
%    \end{cases} \, ,
%\end{equation}
%\end{linenomath*}
where $\sigma_i$ is the 1$\sigma$ measurement error at the observation time $t_i$. The nonzero off-diagonal terms come from a correlated red noise, where $\sigma^2$ and $\tau$ are the variance and the correlation time of the variability. The null hypothesis is a flat mean amplitude with a correlated red noise, equivalent to a pure DRW model. We consider three periodic models. These include a sinusoidal model as well as two fiducial bursty accretion models, assuming mass ratios q=0.11 and 0.43. We adopt a maximum likelihood approach for the parameter fitting and model comparison. The likelihood function is give by 
\begin{linenomath*}
\begin{equation}
\label{eq:likelihood_func}
L \propto
\det | C |^{- \frac{1}{2}}
\exp \bigg[ - \frac{1}{2} (X_i - M_i) \left( C^{-1} \right)_{ij} (X_j - M_j) \bigg] \, ,
\end{equation}
\end{linenomath*}
where $X_i$ is the observed flux and $M_i$ the model flux at the observation time $t_i$. 

First, we test if the $q$=0.11 model could explain our detected periodicity by maximizing the likelihood function without any limitation on the parameters. We use the \textit{emcee} package to determine the best-fit parameters and their uncertainties. We initiate 100 individual chains to sample the maximum likelihood function for 500 steps. Then, we remove the first 250 steps as a burn-in process. The 1$\sigma$ error is determined by the remaining 250 steps from 100 chains at the 84.14 and 15.86 percentiles. The best-fit q=0.11 bursty model period along with the 1$\sigma$ error are listed in Table \ref{tab:measurement}, consistent with the periodicity found in the periodogram analysis. 

%%We take the $g$-band best-fit value as the final estimate for the period, and adopt the ranges encompassing all the $griz$ 1$\sigma$ errors as the final 1$\sigma$ error for the period (i.e., 1597$^{+77}_{-121}$ days). 

Then, we compare three models (sinusoidal + red noise, circumbinary accretion + red noise, and a pure stochastic red noise) using maximum likelihood estimation. All the calculations are done in flux units. In a single-band fit, the sinusoidal model has six free parameters: red noise amplitude, red noise correlation time, period, phase, amplitude and average magnitude. The more bursty, circumbinary disk accretion variability model also has six free parameters: red noise amplitude, red noise correlation time, period, phase shift, amplitude of variation, and the magnitude zero point. A DRW model has three free parameters: red noise amplitude, red noise correlation time, and mean magnitude. 
%During the model comparison, we allowed the periodicity to vary within the window of 1437 to 1693 days. 
%Under the assumption that the periodicity is the same across four bands, we fit the $g$-band light curve first to determine the best-fit period {\color{green} and phase} for each of the three candidate models. We then adopt the the best-fit $g$-band period {\color{green} and phase} to the other three bands. As a result, the number of free parameters reduces from {\color{green} six} to {\color{green} four} for the $riz$ bands. 

We also do a combined fit making use of the light curves from all four bands. To help break parameter degeneracy, the periodicity, phase and red noise correlation timescale are fixed to be the same across different bands. In a combined fit with the periodic models, we have fifteen free parameters, including the mean flux, model amplitude and red noise amplitude in each band, as well as the periodicity, the phase and the red noise correlation timescale which are the same across different bands. For the pure DRW model, there are nine model parameters, including the mean flux and the red noise amplitude in each band, and a red noise correlation time which is the same across different bands.

To compare different models, we adopt the Bayesian information criterion (BIC), which is defined as
\begin{linenomath*}
\begin{equation}
{\rm BIC} = -2 \ln (L) + k \ln (N) \, ,
\label{eq:BIC}
\end{equation}
\end{linenomath*}
%where $k$ is the number of free model parameters. A lower BIC indicates the more preferred model. We list the BIC value differences between models in Table 1. In each band, the $q$=0.11 accretion model always has the smallest BIC values among all the three periodic models considered. We thus conclude that the $q$=0.11 accretion model is the best model for the observed light curves.  
where $k$ is the number of free model parameters and $N$ the number of data points. A lower BIC value indicates the more preferred model. We adopt the proportional constant to be unity in Eq (\ref{eq:likelihood_func}).

%%%%%%%%%%%%%%%%%%%%%%%%%%%%%%%%%%%%%%%%%%%%%%%%%%%%%%%%%%%%%%%%
%%%%%%%%%%%%%%%%%%%%%%%%%%%%%%%%%%%%%%%%%%%%%%%%%%%%%%%%%%%%%%%%
%%%%%%%%%%%%%%%%%%%%%%%%%%%%%%%%%%%%%%%%%%%%%%%%%%%%%%%%%%%%%%%%
% data and measurement tables:
%
%
%\newgeometry{margin=1cm}
%\begin{landscape}
\begin{table*}
\centering
\begin{tabular}{lccccc}
\hline\hline
~~~~~~~~~~Parameter~~~~~~~~~~~~~~~~~~~~~~~~~~~~~~~~~~~~~~~~~~~~ & g & r & i & z & griz  \\
\hline
$P_{{\rm GLS}}$ (days)\dotfill(1) & 1607$\pm$7 & 1615$\pm$9 & 1632$\pm$8 & 1607$\pm$10 & --  \\
$P_{{\rm Acc,\,q=0.11}}$ (days)\dotfill(2) & 1511$^{+34}_{-55}$  & 1466$^{+64}_{-12}$ & 1506$^{+128}_{-61}$ & 1562$^{+248}_{-99}$ & 1476$^{+128}_{-5}$\\
%BIC$_{{\rm Acc,\,q=0.11}}$$-$BIC$_{{\rm sin}}$\dotfill(3) & $-$683 & $-$632 & $-$513 & $-$111 & $-$189 \\
%BIC$_{{\rm Acc,\,q=0.43}}$$-$BIC$_{{\rm sin}}$\dotfill(4) & $-$538 & $-$342 & 456 & 437 & 1391 \\
BIC$_{{\rm Acc,\,q=0.11}}$$-$BIC$_{{\rm DRW}}$\dotfill(3) & $-$19.7 & $-$23.7 & $-$13.3 & $-$13.7 & $-$96.7 \\
BIC$_{{\rm Acc,\,q=0.43}}$$-$BIC$_{{\rm DRW}}$\dotfill(4) & $-$9.7 & $-$7.0 & $-$2.5 & $-$1.9 & $-$78.3 \\
BIC$_{{\rm sin}}$$-$BIC$_{{\rm DRW}}$\dotfill(5) & $-$5.5 & $-$5.7 & $+$0.9 & $-$2.9 & $-$47.2  \\
$N$ \dotfill(6) & 212 & 223 & 222 & 227 & 884 \\
%$\beta \equiv dln(F_{\lambda})/dln(\lambda)$\dotfill(6) & $-$1.68$\pm$0.40 & $-$0.75$\pm$0.34 & $-$1.80$\pm$0.34 & $-$2.41$\pm$1.01 \\ removed since beta and alpha are redundant
$\alpha \equiv dln(F_{\nu})/dln(\nu)$\dotfill(7) & $-$0.32$\pm$0.40 & $-$1.25$\pm$0.34 & $-$0.20$\pm$0.34 & 0.41$\pm$1.01 & -- \\
$A_{{\rm obs}}$ (mag)\dotfill(8) & 0.229$\pm$0.003 & 0.162$\pm$0.002 & 0.162$\pm$0.002 & 0.157$\pm$0.004 & --  \\
$\tau_{\rm DRW}$ (days) \dotfill(9) & 653 & 716 & 629 & 849 & 701 \\
{$k_{\rm DRW}$ / $k_{\rm q=0.11}$ / $k_{\rm q=0.43}$ / $k_{\rm sin}$ \dotfill(10)} & {3/6/6/6} & {3/6/6/6} & {3/6/6/6} & {3/6/6/6} & {9/15/15/15} \\
%{\color{green}$P_{\rm ACF}$ (days) \dotfill(10)} & {\color{green} 1514$^{+8}_{-5}$} & {\color{green} 1565$^{+8}_{-4}$} & {\color{green} 1508$^{+5}_{-5}$} & {\color{green} 1575$^{+11}_{-10}$} & {\color{green}--} \\
\hline
\end{tabular}
\caption{\textbf{Measurements of \obj .} 
Line (1): Period and 1$\sigma$ error (estimated from bootstrap re-sampling) from the generalized Lomb-Scargle (GLS) periodogram.
Line (2): Best-fit period and 1$\sigma$ error (statistical) from MCMC fitting the $q$=0.11 accretion model independently in different bands assuming a correlated red noise.
%
%Lines (3) \& (4): Bayesian information criterion (BIC) differences between models. A lower BIC value indicates the more preferred model. The periodic models considered include a sinusoidal model (expected for relativistic Doppler boost) and two bursty, circumbinary accretion models assuming q=0.11 and 0.43. %  
Lines (3)--(5): Bayesian information criterion (BIC) differences between a periodic model and the null hypothsis, i.e., stochastic quasar variability characterized by a damped random walk (DRW) model. The periodic models considered include two bursty, circumbinary accretion models assuming $q$=0.11 and 0.43, and a sinusoidal model (expected for relativistic Doppler boost). A negative $\Delta$BIC indicates that a periodic model is more preferred over a pure stochastic variability. $\Delta$BIC$<$$-$10 suggests strong evidence.  
%In all periodic models we have fixed the $riz$ period (i.e., corresponding to the frequency with the largest periodogram power) to be the best-fit $g$-band value to avoid parameter degeneracy issues given the limited data points.  
Line (6): Number of data points.
Line (7): Power-law index of the continuum from spectral modeling. Errors represent 1$\sigma$ uncertainties generated from 100 Monte Carlo simulations. 
Line (8): Variability amplitude from the best-fit sinusoidal model. Errors represent 1$\sigma$ statistical uncertainties. 
Line (9): Best-fit correlation time in the DRW model.
{Line (10): Number of free parameters for each of the model.}
%Line (10): Best-fit period and 1$\sigma$ error (statistical) from MCMC fitting the Aauto-correlation function.
%Lines (7)--(9): the observed variability amplitude ratio between any given two bands. Uncertainties are 1$\sigma$ from error propagation.
}
\label{tab:measurement}
\end{table*}
%\end{landscape}
%\restoregeometry

Table \ref{tab:measurement} lists the BIC from our MCMC analysis for the model fitting and comparisons. Three periodic models are compared against the null hypothesis of a pure stochastic variability. A lower BIC value indicates the more preferred model, and a BIC difference of ${<}-$10 suggests strong evidence. In each band, the $q=0.11$ accretion model always has a negative BIC difference (i.e., suggesting that it is more preferred than a pure stochastic variability), which is also the smallest among all the three periodic models considered. The observed light curves of \obj\ statistically prefer the $q$=0.11 accretion model over the other models in all four bands. Taking the fit that combines all bands for example, the BIC difference between the $q$=0.11 accretion model and the pure stochastic quasar variability model translates to a likelihood ratio of, at least, exp[($-$96.7)/($-$2)]${\sim}10^{21}$ (Eq (\ref{eq:BIC})). We thus conclude that the $q=0.11$ accretion model to be the best model for the observed light curves.

We have tested that our qualitative conclusion (that the $q=0.11$ accretion model is preferred over a sinusoidal model from having smaller BIC values) still holds assuming a background of pure white (flat spectrum) noise instead (i.e., with zero off-diagonal terms in the covariance matrix). We show the best-fit $q=0.11$ accretion models under white noise (thin solid curves) in Figure \ref{fig:lc} for illustration purposes only. We have also tested an eccentric Doppler boost model, but the $q=0.11$ circumbinary accretion model still has the lowest BIC.

%Taking the $g$ band for example, the BIC difference between the $q$=0.11 bursty accretion model and the sinusoidal model (i.e., $\Delta$BIC = $-$683) translates approximately to a likelihood ratio of exp($-$683/2)${\sim}???$ Assuming Gaussian distributions, 

%We estimate the Eddington ratio as $L_{{\rm bol}}/L_{{\rm Edd}}$, where the Eddington luminosity is $L_{{\rm Edd}}=M_{\bullet}/M_{\odot}\times10^{38}$ erg s$^{-1}$ and the bolometric luminosity $L_{{\rm bol}}$ is calculated from $L_{{\rm 1350}}$ using the bolometric correction BC$_{{\rm 1350}}$=3.81 from the composite quasar spectral energy distribution of \citep{Richards2006}.

\subsection{Relativistic Doppler Boost Modeling} 

The multi-band light curves enable us to conduct an independent, quantitative test of the relativistic Doppler boost hypothesis. The relativistic Doppler boost predicts unique and robust frequency-dependent variability amplitudes in different bands that can be tested with multi-color data \citep{DOrazio2015a}. We adopt the total mass of the hypothesized binary in \obj\ as $M = 10^{8.4\pm0.1} M_{\odot}$ (1$\sigma$ statistical error) assuming the virial black hole mass estimated from \MgII . We measure the spectral indices of the continuum by fitting broken power law models over four wavelength windows corresponding to the $griz$ bands. Table \ref{tab:measurement} lists the resulting broken power-law indices. 

For a binary in a relativistic circular orbit, the observed frequency of the emitted photons from the secondary's accretion disk will change due to the relativistic motion \citep{DOrazio2015a,Charisi2018}. The number of photons 
$N$, where $N \propto {\cal F}_\nu / \nu^3$, with ${\cal F}_\nu$ being the flux at a specific frequency $\nu$, is Lorentz invariant. 
The photons are Doppler-boosted by a factor 
\begin{linenomath*}
\begin{equation}
{\cal D} = \frac{1}{\gamma \left(1 - v_\parallel / c\right)} \ ,
\end{equation}
\end{linenomath*}
where $v$ is the orbital velocity, $v_\parallel$ is the line-of-sight component, and $\gamma = \left(1-\beta^2\right)^{-1/2}$. Assuming that the emitted radiation has a power-law spectrum ${\cal F}_\nu \propto \nu^{\alpha_\nu}$, the observed flux is
\begin{linenomath*}
\begin{equation}
{\cal F}_\nu^{\rm obs} = {\cal D}^{3-\alpha_\nu} {\cal F}_\nu^{\rm em} \ .
\end{equation}
\end{linenomath*}
For a binary in a circular orbit, the Doppler-boost variability to first order in $\beta$ is
\begin{linenomath*}
\begin{equation}
\frac{\Delta {\cal F}_\nu}{{\cal F}_\nu} = \left(3-\alpha_\nu\right) \beta \cos\phi \sin i \ ,
\end{equation}
\end{linenomath*}
where $v$ is the orbital velocity of the more luminous black hole (assumed to be the less massive secondary black hole, whereas the primary black hole is assumed to contribute negligible flux), $i$ is the inclination of the binary orbit with respect to the line-of-sight (defined such that $i$=90 degrees for an edge-on view and 0 degrees for a face-on view), and $0 \le \phi \le 2\pi$ is the phase of the orbit. We take the orbital separation to be effectively constant over the course of the observation since the time elapsed in the rest frame is much less than the coalescence timescale of the binary. Taking the $g$ band for example, the amplitude of the variability is 0.229 mag (Table \ref{tab:measurement}), corresponding to $\Delta F_{\nu}/F_{\nu}\sim \pm$0.229. To explain this, a line-of-sight velocity amplitude of $vsin(i)\approx$0.069$c$ would be needed, considering the $g$-band power-law index $\alpha_g{\sim}$$-$0.32 (Table \ref{tab:measurement}). 

%For the four bands, we have $\alpha_g=-0.32\pm0.40$, $\alpha_r=-1.25\pm0.34$, $\alpha_i=-0.20\pm0.34$, and $\alpha_z=0.42\pm1.01$. 

%{\color{red} Miguel: See Figure 1 of DOrazio2015. We should show the parameter space in mass, inclination, and mass ratio instead. We can assume that $f_2=1.0$, but comment that this is an extreme assumption and effectively the allow parameter space is zero if $f_2<0.8$ even if mass is significanly underestimated.}  

We calculate the frequency-dependent variability amplitude ratios expected from relativistic Doppler boost (i.e., relativistic beaming, or RB for short) to compare with the observations. Taking the $gr$ bands for example, the RB model predicts $A_{g, {\rm RB}}/A_{r, {\rm RB}}=(3-\alpha_{g})/(3-\alpha_{r})=0.78\pm0.11$ (1$\sigma$), where $\alpha\equiv dln(F_{\nu})/dln(\nu)$. The observed $A_{g, {\rm obs}}/A_{r, {\rm obs}}$ is $1.41\pm0.03$. The RB hypothesis is therefore ruled out at $\gtrsim$5$\sigma$.

\begin{figure}
\centering
\includegraphics[width=0.5\textwidth]{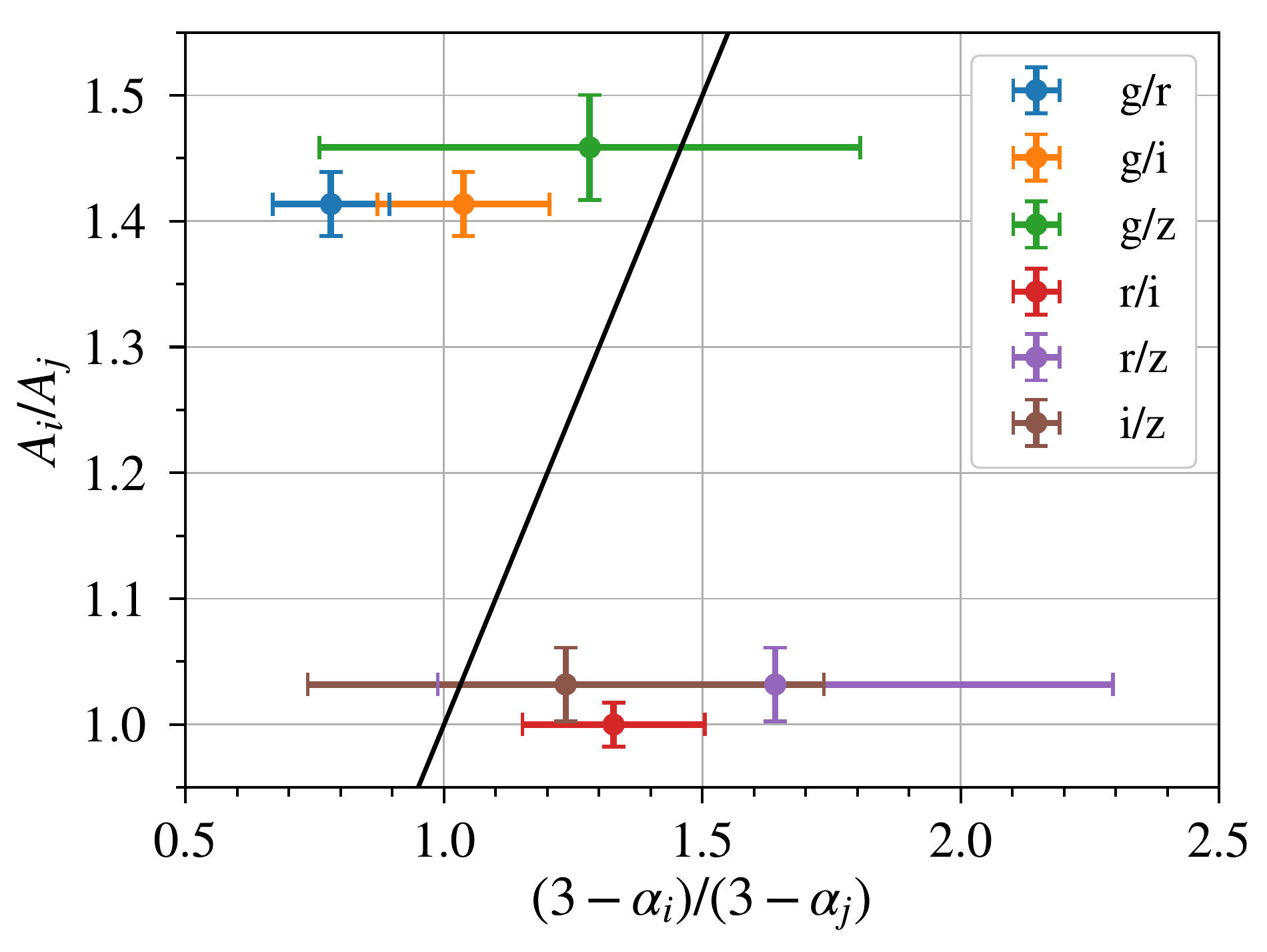}
\caption{Observed frequency dependent variability amplitude ratio for each band combination compared with the expected values from relativistic Doppler boost. The black line represents the 1 to 1 relation. Error bars denote 1$\sigma$ uncertainties.
}
\label{fig:rb_amplitude_ratio_test}
\end{figure}
\begin{figure*}
\centering
\includegraphics[width=0.35\textwidth]{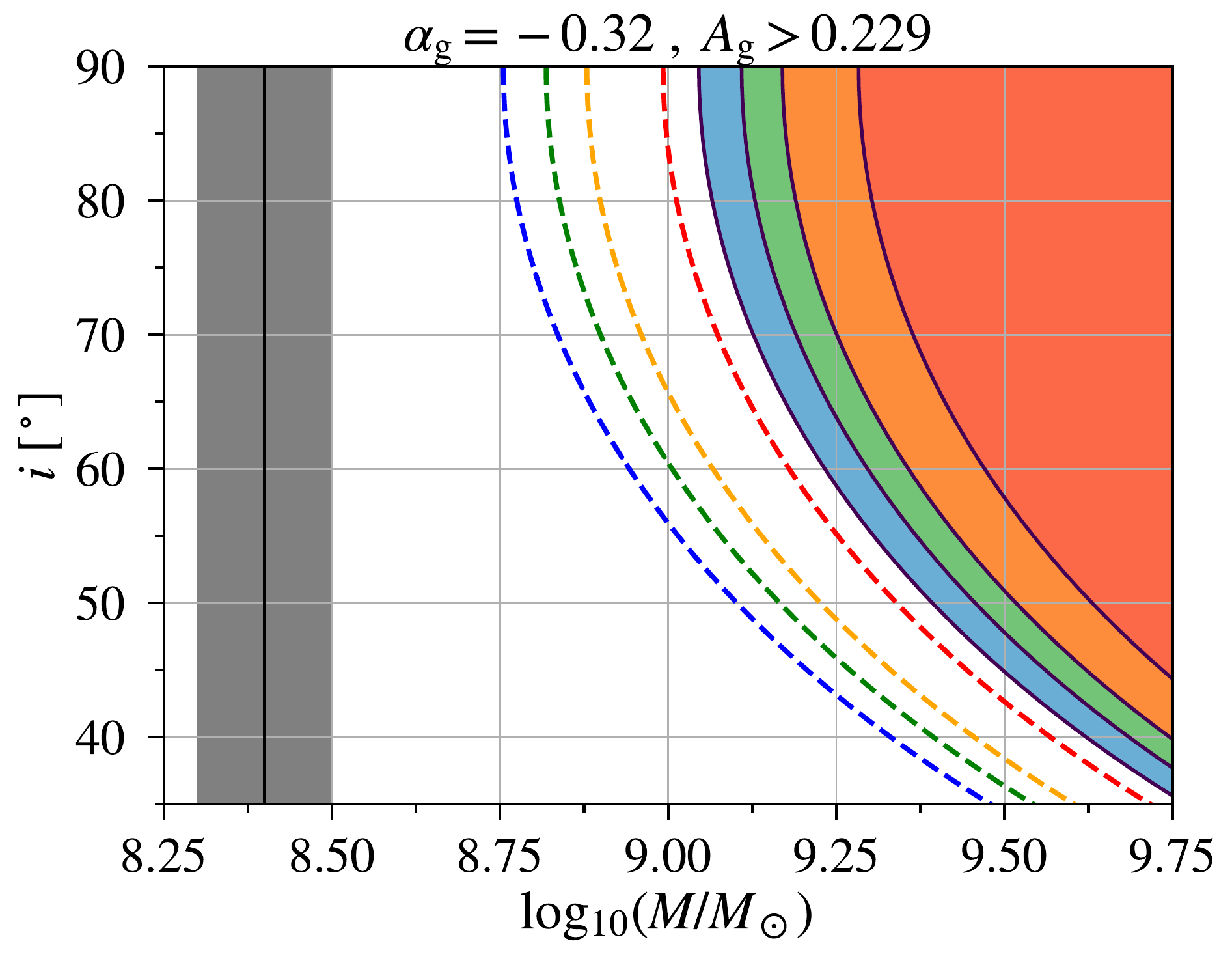}
\includegraphics[width=0.35\textwidth]{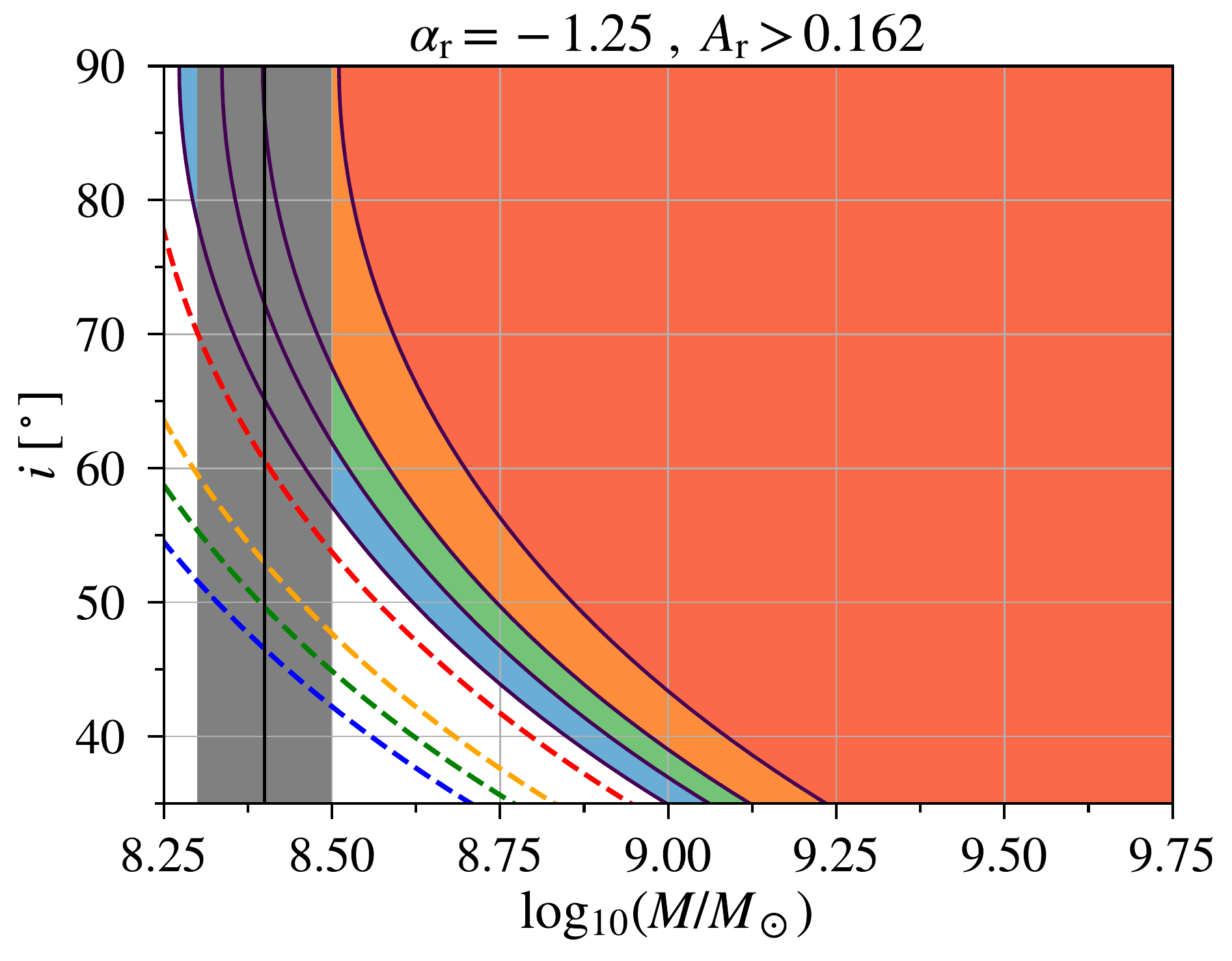}
\includegraphics[width=0.35\textwidth]{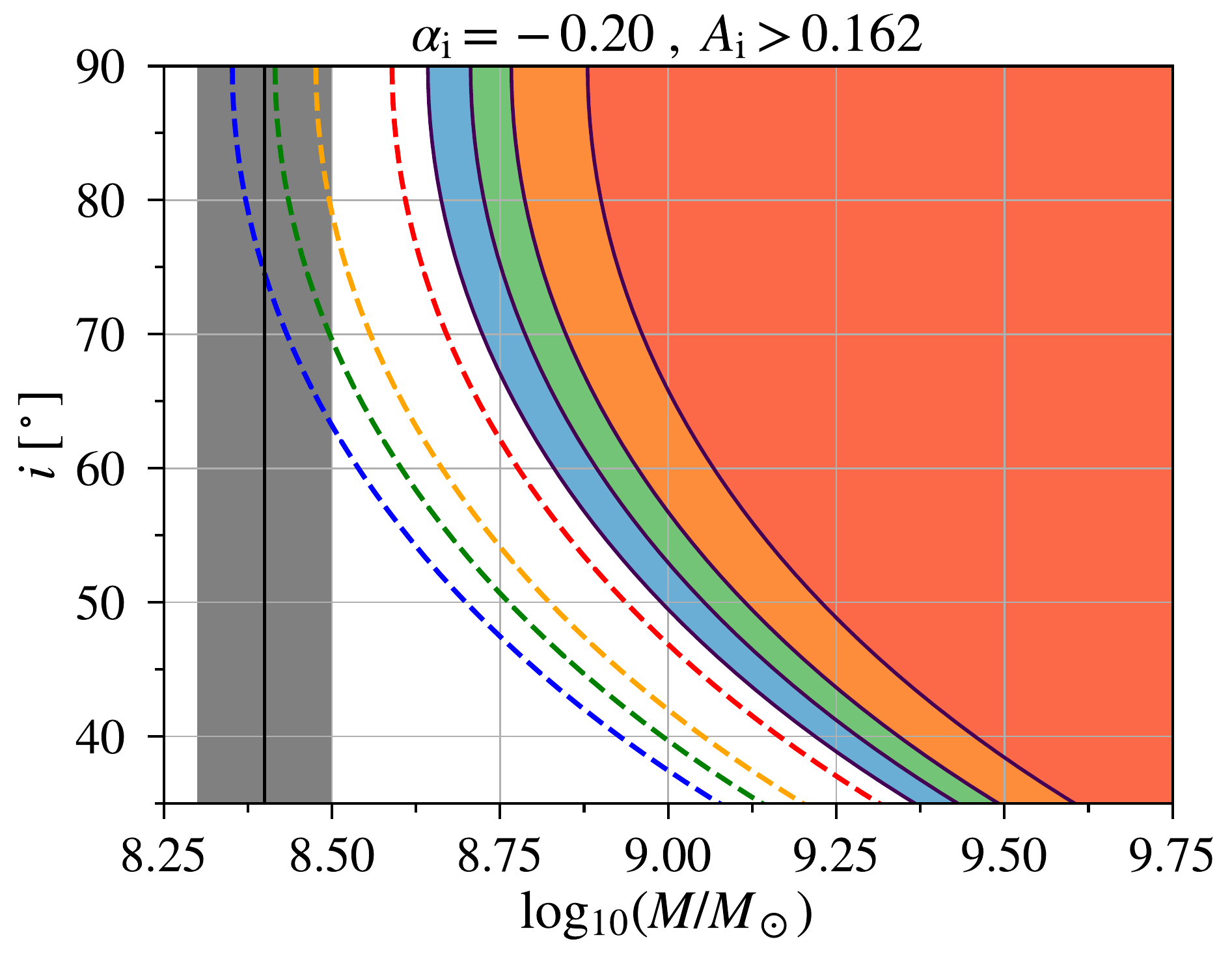}
\includegraphics[width=0.35\textwidth]{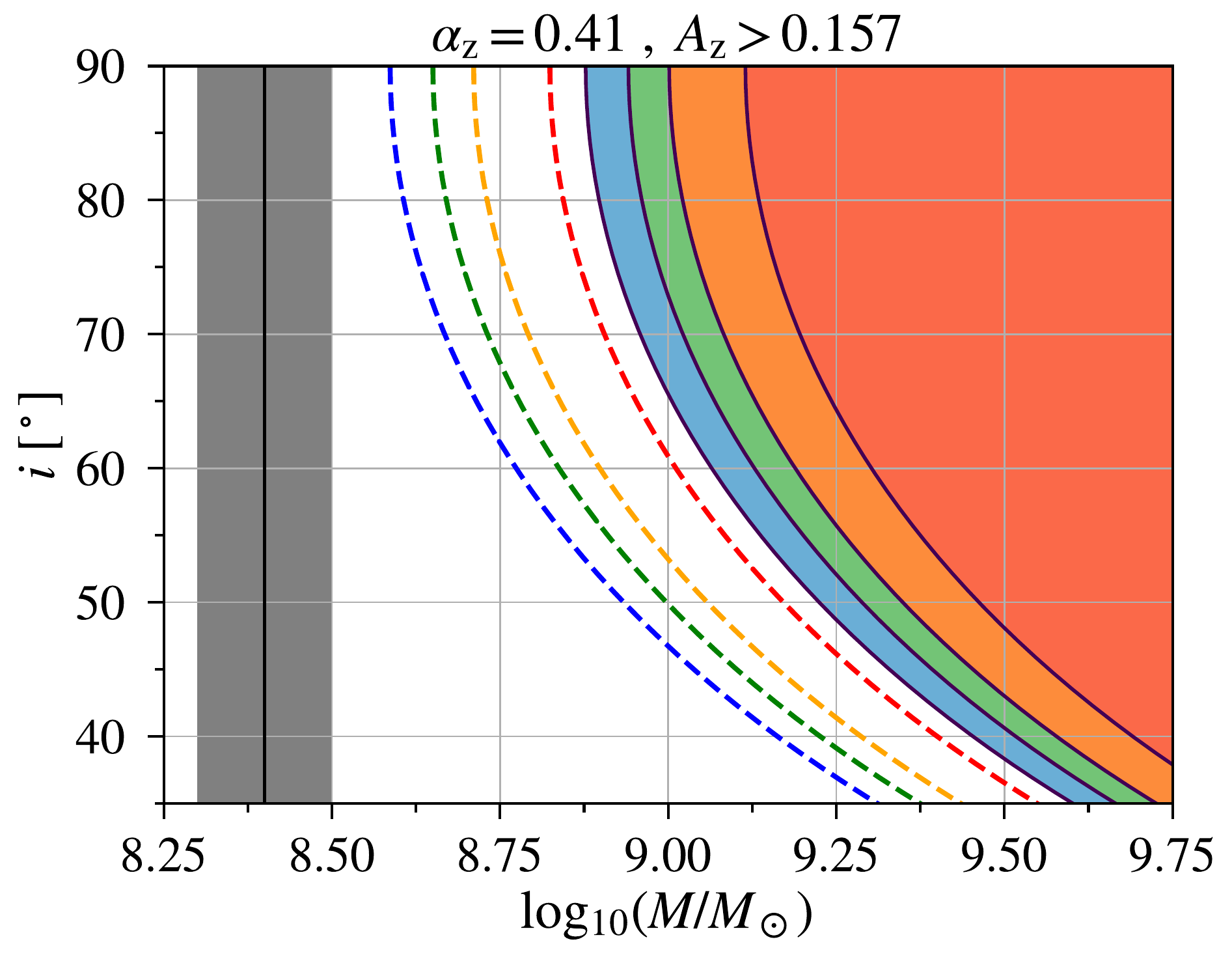}
\caption{Parameter space estimates for the relativistic Doppler boost model.
The four panels represent $griz$ bands. In each panel, the dashed contours represent $f_2=1.0$ whereas the shaded contours denote $f_2=0.8$, where $f_2$ is the fraction of the total emission coming from the secondary black hole. Different colors show different mass ratios with $q$=0.0, 0.05, 0.1, and 0.2 for blue, green, orange, and red, respectively. The vertical solid line with gray shades indicate our virial mass estimate and its 1$\sigma$ statistical error for the total black hole mass. The orbital inclination angle $i$=90 degrees for an edge-on view, and 0 degree for a face-on view.
}
\label{fig:rb_amplitude_test}
\end{figure*}

Figure \ref{fig:rb_amplitude_ratio_test} shows the observed variability amplitude ratio ($A_i/A_j$ where $i$ and $j$ represent two bands) compared with the expected value inferred from relativistic beaming (RB) for each band pair, which is $(3-\alpha_i)/(3-\alpha_j)$. The RB model is being ruled out at $\gtrsim$5$\sigma$ considering the $gr$ bands and at $\sim$2$\sigma$ for the $gi$ and $ri$ bands.

Figure \ref{fig:rb_amplitude_test} shows the parameter space that allows for a flux variability greater than a fiducial value of 16\%--23\% in order to explain the observed values (Table \ref{tab:measurement}). The parameters considered are the total black hole mass $M$, mass ratio $q$, orbital inclination $i$, and the fraction of the total emission coming from the secondary black hole $f_2$ \citep{DOrazio2015a}. Our other fiducial model parameters are $P_{{\rm orb}}$ = 1.7 yr, and $\alpha$ = $-$0.32, $-$1.25, $-$0.20, and 0.41 in the $griz$ bands (Table \ref{tab:measurement}). There is little to no parameter space for the RB hypothesis to work, because the required total black hole mass would be too large to reproduce the observed, large variability amplitudes in \obj , unless all the following three requirements are met: 1. the total black hole mass is significantly underestimated by the virial estimate, even when accounting for a 0.5-dex systematic uncertainty \citep{Shen2013}, 2. $>$80\% of the optical light is contributed by emission from a mini accretion disk fueling the secondary black hole, and 3. the system is viewed close to being edge-on. 

Our estimates on the periodic variability amplitudes (Line 8 in Table \ref{tab:measurement}) do not include contribution from a stochastic background of red noise; accounting for all the observed variability amplitudes instead would make the tension even stronger.

%\noindent \textbf{Periodic self-lensing to constrain the inclination angle?}
%
%
%\noindent \textbf{Probability of galaxy hosting a BSBH in the PTA band.} Pulsar timing arrays are sensitive to gravitational waves from supermassive black-hole binaries with periods of order years. 
%The probability of a galaxy hosting a BSBH in the PTA band is
%
%\begin{equation}
%{\cal P} = \frac{\tau_{\rm c}}{\tau_{\rm z}}\int_{0.25}^{1} \frac{{\rm d}N}{{\rm d}t} \left(M_*, z ,\mu_*\right) \tau_z \ {\rm d}\mu_* \ .
%\end{equation}
%

\subsection{Gravitational-wave implications and prospects.} 

\begin{figure}
\centering
\includegraphics[width=0.5\textwidth]{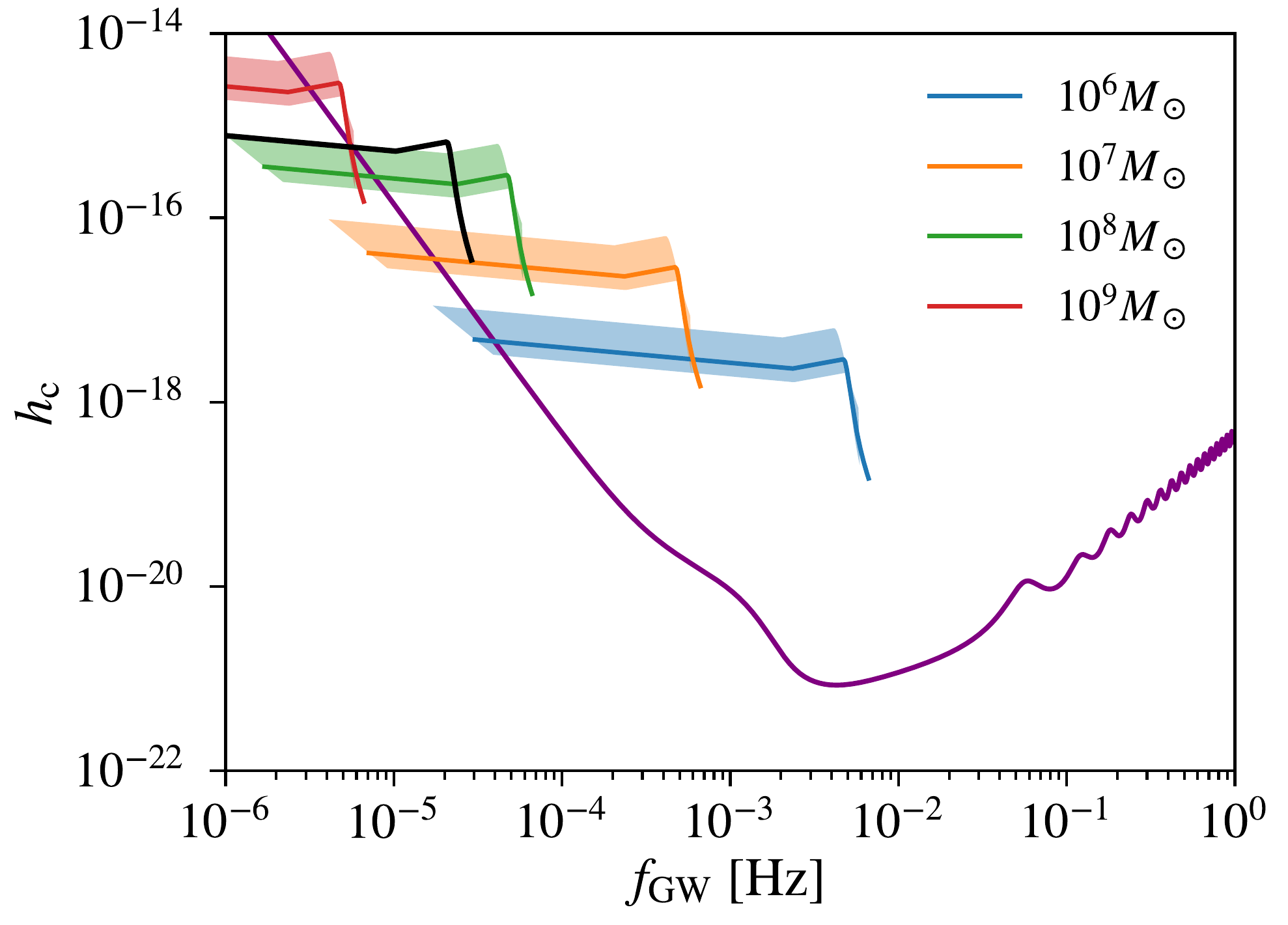}
\caption{Prospect for LISA detection of a source similar to the candidate BSBH in \obj\ but 5 years before coalescence.
The purple curve represents the expected LISA sensitivity limit assuming a 5-yr observation \citep{Robson2019}. The black curve denotes the gravitational-wave signal of a BSBH at $z = 1.53$ with mass $10^{8.4} M_\odot$ and mass ratio $q = 0.1$ beginning at 5 yrs before coalescence, i.e., from the inspiral phase (low frequency) to the final merger and ringdown (high frequency). The blue, orange, green, and red shaded regions correspond to mergers with a primary mass of $10^6 M_\odot$, $10^7 M_\odot$, $10^8 M_\odot$, and $10^9 M_\odot$, respectively, at the same redshift with mass ratios ranging within $0.05 \le q \le 0.5$. The blue, orange, green, and red lines correspond to $q = 0.1$. 
}
\label{fig:lisa}
\end{figure}

The GW strain amplitude of a circular binary in the quadrupolar approximation is
\begin{linenomath*}
\begin{equation}
h_0 = \frac{4 G^{5/3}}{c^4} \frac{\mu M^{2/3} \omega^{2/3}}{D_{\rm L}} \ ,
\end{equation}
\end{linenomath*}
where $\mu = Mq/(1+q)^2$ is the reduced mass, $M$ is the total mass, $\omega = 2 \pi f_{\rm orb}$, and $D_{\rm L}$ is the luminosity distance to the source. From our parameter estimation, the inferred strain amplitude is $h_0 \sim 9.8{\times}10^{-19}$, which makes this binary effectively undetectable by current Pulsar Timing Arrays (PTAs) as an individual source \citep{Zhu2014}. Recent PTA upper limits on the stochastic background have been used to constrain the ensemble properties of BSBH candidates \citep{Sesana2018,Holgado2018}. A growing census of milli-pc BSBH candidates will be further constrained as the PTA sensitivity improves over time. LISA would be sensitive to BSBH mergers at these given masses and mass ratios. We estimate the SNR from the latest LISA sensitivity curve \citep{Robson2019}. Figure \ref{fig:lisa} shows that the BSBH candidate would eventually merge in the LISA frequency band and a merging binary with the same mass and mass ratio would be detectable during a 5-yr observation with a nominal SNR ${\sim}$15 at redshift ${\sim}$1.5.

%from Miguel: since my preliminary calculations suggested that this binary merges within a fraction of a Myr, I decided to check if this binary would merge in the LISA frequency band and if this binary would be detectable during a 5-yr observation. It turns out that the nominal SNR for the merger of this binary would be SNR~15 at redshift ~1.53 (figure attached)! 

%
%%

\subsection{Alternative Interpretations} 

Unlike the two previously best known BSBH candidates OJ287 and PG1302, \obj\ is not a blazar, nor is its optical emission dominated by contribution from a radio jet, and therefore jet precession cannot explain the periodicity. Precession of a warped accretion disk is unlikely because the amount of obscured continuum emission required would be too large to explain the observed variability amplitude in \obj\ and that the effect is geometrical rather than bursty. The periodicity in \obj\ (i.e., rest-frame 1.7 yr) is close to the expected value ($\sim$200 days) inferred from its black hole mass assuming a scaled-up quasar version \citep{King2013} of low-frequency accretion disk quasi-periodic oscillations (QPOs; e.g., from strong resonances in the accretion flow) as seen in the X-ray light curves of X-ray binaries \citep{Vaughan2005a}. However, the characteristic bursty light curves would be difficult to explain with Lense-Thirring precession \citep{Bardeen1975} of a geometrically thick accretion flow near the primary black hole, with which low-frequency QPOs in X-ray binaries are associated \citep{Stella1998,Ingram2009}. QPOs in X-ray binaries show drifts in period, phase, or amplitude \citep{vanderKlis1989}, and future continued monitoring observations can constrain the possibility of an optical QPO.

\subsection{Implication for the Merger Hypothesis and Comparison with Theoretical Event Rates} 

Among a parent sample of {625} quasars we have detected one strong candidate BSBH, whose estimated gravitational-wave inspiral time (0.17 Myr) is about $\sim$10$^2$--10$^3$ times shorter than estimates for quasar lifetimes \citep{Martini2001,Yu2002}. This implies that most quasars could be binary systems with a much larger binary separation that the circumbinary disk does not yet exist, which is unsurprising given the merger hypothesis \citep{volonteri03,hopkins08,shen09,Haiman2009}. Previous work has predicted the event rates of BSBHs that are detectable as periodic quasars \citep{haiman09}. The most recent work by \citet{Kelley2019} combines cosmological hydrodynamic simulations, semi-analytic binary merger models, and analytic quasar spectra and variability prescriptions. Given DES sensitivity (assuming a typical single epoch 5$\sigma$ point source depth of {$\sim$23.5} AB mag), the expected number of detectable periodic quasars from circumbinary disk accretion variability at redshift z$\sim$1.5 with observer-frame periods between 0.5 and 5.0 yr is {$\sim$80} in an all-sky survey ($\sim$30,000 deg$^2$), or $\sim$1 per {380} deg$^2$ (see right panel in Figure 6 of \citealt{Kelley2019}). This is {$\sim$70} times lower than our detection rate\footnote{Our estimated detection rate depends on the depth of the parent spectroscopic quasar sample, which is incomplete. We do not have a complete quasar sample down to DES depth.} of {$\sim$one strong BSBH candidate from circumbinary accretion variability per 5 deg$^2$} at face value. As further discussed by \citet{Chen2020}, this apparent discrepancy is likely explained by the fact that our sample is dominated by less massive quasars at high redshift given our deep survey over a small area. As a result we are effectively measuring the differential detection rate (which is a function of redshift and BH mass) rather than the cumulative detection rate as quoted by \citet{Kelley2019}.

There are still significant uncertainties that prevent a fair comparison between our detection rate and theoretical predictions and PTA limits. First, theoretical event rates are still highly uncertain. The most significant uncertainty is on the inspiraling timescales, which could lead to highly uncertain estimates on the number of detectable binaries in the circumbinary accretion disk phase.

Second, the PTA upper limits are still subject to model uncertainties regarding the evolutionary history of a binary from large to small separations where GW emission dominates. PTA upper limit is model independent only for a particular binary separation range that corresponds to the PTA frequency. To extrapolate this to other separations (i.e., going from PTA frequency to the frequency relevant for periodic quasars), one needs to invoke assumptions on the evolutionary timescales. However, there is still no self-consistent model that can deal with the full evolution considering the effects of gas and stars, and so a high binary fraction at mili-parsec scales may not necessarily be in direct tension with the PTA upper limits. For example, if a binary stalls at large separations, or if it sweeps quickly through the PTA sensitivity range, there would be no PTA signal, even if the true binary fraction were high. 

Finally, PTA is most sensitive to the most massive binaries at low redshift. However, our sample is most sensitive to the $\sim10^8M_{\odot}$ systems at intermediate redshift. A small binary fraction for the most massive black holes at z=0 from the PTA upper limits does not directly translate into the same binary fraction for the less massive black holes at $z\sim1.5$.

\subsection{Comparison with Previous Work}

As further shown in \citet{Chen2020}, our detection rate {of all candidate periodic quasars (not just BSBH candidates from circumbinary accretion variability)}, i.e., {$\sim$0.8\%}, is {$\sim$4--80} times of those from previous searches using other surveys \citep{Graham2015,Charisi2016,Liutt2019}, even though our selection criteria are more stringent. For example, we request $>$ three cycles in at least two bands, whereas only 1.5 cycles were adopted by \citet{Graham2015} and only one-band data were available. This is not a fair comparison, however, because {our sample is probing less massive quasars at higher redshifts than those studies in previous shallower surveys over larger areas. As suggested in \citet{Chen2020}, the significantly higher detection rate of periodic quasars found in our sample may be interpreted as the redshift evolution of the fraction of BSBHs, i.e., the binary fraction is larger at higher redshifts at a fixed BH mass.} 

In addition, previous datasets lacked the long time baseline and/or sensitivity to discover similar systems as \obj . Given shorter time baselines and lower sensitivities, false positives and/or false negatives would have been more likely to significantly bias the apparent detection rates because of stochastic background variability. In particular, \citet{Liutt2019} has rejected most of the candidates found in their previous searches \citep{Liutt2015,Liutt2016} by continued monitoring of the ``best candidates''. While this demonstrates the importance of a long time baseline in rejecting false positives due to stochastic background variability, it does not address the question of possibly missing false negatives in those that have not been continuously monitored. A long time baseline for the full parent sample (i.e., not just the ``best candidates'' selected based on short-baseline light curves) is needed to robustly quantify the true detection rate.  

In summary, the quality of the data (i.e., long time baseline, high sensitivity) is more important than the quantity of the data (i.e., size of the parent quasar sample) because the systematic error (e.g., bias from false positives and/or false negatives caused by stochastic background variability) is likely to be larger than the statistical error. Even though we have a much smaller sample of quasars in the parent sample, our detection rate is still likely to be more reliable than those from previous work based on shorter and shallower surveys of larger areas.

\section{Conclusion and Future Work}\label{sec:sum}

Our results on \obj\ may provide the first, strong evidence for circumbinary accretion variability as the physical origin for periodic quasar (optical) light curves. Sensitive, long-term, multi-color light curves are key in disfavoring the competing relativistic-Doppler-boost hypothesis for \obj . Relativistic Doppler boost has been previously shown to best explain the characteristic periodic optical light curves and UV observations of PG1302$-$102 \citep{DOrazio2015a}. We speculate that various mechanisms may be at work in different systems, such that the case for PG1302$-$102 may not necessarily apply to \obj\ or other periodic quasars. 

Recently, using a combination of cosmological, hydrodynamic simulations, comprehensive semi-analytic binary merger models, and analytic active galactic nucleus spectra and variability prescriptions, \citet{Kelley2019} suggests that hydrodynamic variability should be $\sim$5--25 times more common than relativistic Doppler boost in producing periodic quasar light curves in synoptic surveys. Our result suggests that hydrodynamic circumbinary accretion variability may indeed be a viable option to explain periodic light curves, at least for some, if not most, quasars as BSBH candidates, although we cannot draw large inferences from just a single detection. Alternatively, precession of a radio jet is likely ruled out, because unlike OJ287 \citep{valtonen08}, \obj\ is not a blazar (with a 3$\sigma$ radio flux density upper limit of $<$0.5 mJy at 1.4 GHz and $<$0.4 mJy at 3 GHz), nor is its optical emission dominated by contribution from a radio jet.

While we have adopted the simulated light curves of \citet{Farris2014} as the baseline model, our conclusion is not sensitive to this particular choice because similar characteristic bursty light curves are seen in other independent simulations of circumbinary accretion disks around BSBHs \citep[e.g.,][]{MacFadyen2008,Shi2012,Roedig2012,DOrazio2013,Gold2014a,Shi2015,Tang2018}. While the archival SDSS data has been necessary in extending the time baseline for a statistically significant periodicity detection, the light curve was only well sampled by the new DES observations in terms of sensitivity and cadence, and there were significant observational gaps. The existing data cannot definitively discriminate between the $q$=0.11 and $q$=0.43 circumbinary accretion variability models, although $q$=0.11 is tentatively preferred (Table \ref{tab:measurement}). We consider these two $q$ values as baseline examples because they represent two characteristic regimes in the light-curve behaviors (Figure 9 of \citealt{Farris2014}). In both regimes there is a strong peak in the periodograms of the simulation-predicted light curves corresponding to the orbital frequency of the overdense lump. Adopting a mass ratio of $q$=0.11, $t_{{\rm orb}} = t_{{\rm period}}$ (whereas $t_{{\rm orb}} \approx 0.2 t_{{\rm period}}$ for $q$=0.43 instead; \citealt{Farris2014}), the inferred binary separation is $d\sim4.4$ milli-parsec (i.e., 5.1 light days, or $\sim$200 Schwarzschild radii), assuming a circular orbit. So the confirmation of this candidate would imply that the system has passed the ``final-parsec'' barrier \citep{begelman80} at a redshift of z=1.53. 

The inferred gravitational-wave inspiral time $t_{{\rm gw}}$ with the preferred system parameters is $\sim$0.17 Myr. This implies that the candidate binary is efficiently emitting gravitational waves and will merge well within the age of the universe, even if environmental effects are neglected. BSBHs with masses of $\sim10^8$--$10^9M_{\odot}$ at redshift $z\gtrsim1$ are generally expected around the time of pre-decoupling \citep{Kocsis2011a}, i.e., when $t_{{\rm gw}}>t_{{\rm visc}}$, where $t_{{\rm visc}}$ is the viscous timescale of the accretion disk. The gravitational-wave strain amplitude is ${\sim}10^{-18}$ at $\sim$37 nHz, which, as an individual source, is ${\sim}10^5$ below the current best sensitivity limit of pulsar timing arrays  \citep{Arzoumanian2018a} to continuous-wave sources, and will also be below the expected SKA sensitivity \citep{Wang2017c}. Laser Interferometer Space Antenna \citep{Amaro-Seoane2017} would be able to detect a source similar to \obj\ but $\sim$5 years before coalescence at $\gtrsim$0.01 mHz with a signal-to-noise ratio of $\sim$15 at redshift 1.5 (Figure \ref{fig:lisa}). 

%future work:

Future sensitive, continued multi-band follow-up imaging is needed to further constrain the significance and nature of the optical light-curve periodicity observed in \obj . While the existing data spans 4.6 cycles, only $\sim$3 are well sampled in multiple bands. There is a $\sim$0.1\% probability that the periodogram peak is caused by stochastic quasar variability (i.e., red noise). The significance of a real periodicity should increase as more cycles are covered. Continuous, sensitive follow up with the Blanco 4m/DECam is on-going to better characterize the light curve properties. Hydrodynamic simulations of circumbinary accretion disks predict additional, weaker peaks in the light curve periodograms at different characteristic frequencies depending on the mass ratio, with many associated harnomics for $q{>}$0.43  \citep{Farris2014}. Future observations may be able to better distinguish between the $q$=0.11 and the $q$=0.43 models (e.g., by searching for evidence for additional weaker peaks in the periodogram and quantifying their characteristic relationships with the primary peak \citep{Charisi2015}). 

The observed SED of \obj\ is similar to normal optically selected quasars that are matched in redshift and luminosity. Future more sensitive UV and/or X-ray observations are needed to put further independent constraints \citep[e.g.,][]{Foord2017} on any potentially characteristic SED features to compare with predictions from circumbinary accretion disk simulations \citep[e.g.,][]{Roedig2012,Tang2018}. While the broad-line region is expected to be well outside the radius of the binary, the circular velocity is about 0.05c, which is much greater than the width of the broad emission lines. Any emission lines originating from the disk could in principle show such shifts, but in practice the broad emission line profile becomes more complex and there are no expected coherent radial velocity drifts in the emission lines with time \citep{shen10}. There could be a shift in the Fe K-$\alpha$ line which probes the inner accretion disk and future sensitive X-ray spectroscopic monitoring is needed to test this \citep{McKernan2015}.  

%The third is that I recommend you say something about circular motion and the possibility of radial velocity shifts. While I appreciate that the BLR is well outside the radius of the binary, the circular velocity is about 0.05c, and much greater than the width of the BLR lines. I think it would be helpful to note that any emission lines originating from the disk could show such shifts, although perhaps they are in the noise compared to the BLR. And have you thought about whether there could be a shift in the Fe Kalpha line?

%broader implications

Our detection of one strong BSBH candidate {due to circumbinary accretion variability} in a sample of {625} spectroscopically confirmed quasars from a {4.6} deg$^{2}$ survey implies a detection rate of $\sim$0.16\%, or 1 per {5} deg$^{2}$, which is {$\sim$70} times higher than the expected event rate \citep{Kelley2019} at face value, although the theoretical rate is still highly uncertain considering unconstrained model assumptions. Our detection rate of candidate periodic quasars in the parent sample is {$\sim$4--80 times} times of those from previous searches using other surveys \citep{Chen2020}, although this is not a fair comparison because previous datasets lacked the long time baseline and/or sensitivity to discover similar systems as \obj . Given shorter time baselines and lower sensitivities, false positives and/or false negatives would have been more likely to significantly bias the apparent detection rates because of stochastic background variability. We have demonstrated using \obj\ that multi-band light curves with high sensitivity and a long time baseline is key to not only identifying periodicity but also sorting out its physical origin. Future large, sensitive synoptic surveys such as the Vera C. Rubin Observatory Legacy Survey of Space and Time \citep{Ivezic2019} may be able to detect hundreds to thousands of BSBH candidates from circumbinary accretion variability.

\section*{Acknowledgements}
We thank the anonymous referee for helpful suggestions that improved the manuscript.
X.L. thanks A. Barth, S. Dodelson, S. Gezari, and A. Palmese for comments, and B. Fields, C. Gammie, K. {G{\"u}ltekin}, Z. Haiman, D. Lai, A. Loeb, D. D'Orazio, S. Tremaine, and X.-J. Zhu for discussions, and LCO director Tod Boroson for granting us DDT observation. W.-T. L. is supported in part by the Gordon and Betty Moore Foundation's Data-Driven Discovery Initiative through Grant GBMF4561 to Matthew Turk and a government scholarship to study aboard from the ministry of education of Taiwan. Y.C.C. and X.L. acknowledge a Center for Advanced Study Beckman fellowship and support from the University of Illinois campus research board. A.M.H. is supported by the DOE NNSA Stewardship Science Graduate Fellowship under grant number DE-NA0003864. Y.S. acknowledges support from the Alfred P. Sloan Foundation and NSF grant AST-1715579. This work makes use of observations from the LCOGT network. 

Funding for the DES Projects has been provided by the U.S. Department of Energy, the U.S. National Science Foundation, the Ministry of Science and Education of Spain, the Science and Technology Facilities Council of the United Kingdom, the Higher Education Funding Council for England, the National Center for Supercomputing Applications at the University of Illinois at Urbana-Champaign, the Kavli Institute of Cosmological Physics at the University of Chicago, the Center for Cosmology and Astro-Particle Physics at the Ohio State University, the Mitchell Institute for Fundamental Physics and Astronomy at Texas A\&M University, Financiadora de Estudos e Projetos, Funda{\c c}{\~a}o Carlos Chagas Filho de Amparo {\`a} Pesquisa do Estado do Rio de Janeiro, Conselho Nacional de Desenvolvimento Cient{\'i}fico e Tecnol{\'o}gico and the Minist{\'e}rio da Ci{\^e}ncia, Tecnologia e Inova{\c c}{\~a}o, the Deutsche Forschungsgemeinschaft and the Collaborating Institutions in the Dark Energy Survey. 

The Collaborating Institutions are Argonne National Laboratory, the University of California at Santa Cruz, the University of Cambridge, Centro de Investigaciones Energ{\'e}ticas, Medioambientales y Tecnol{\'o}gicas-Madrid, the University of Chicago, University College London, the DES-Brazil Consortium, the University of Edinburgh, the Eidgen{\"o}ssische Technische Hochschule (ETH) Z{\"u}rich, Fermi National Accelerator Laboratory, the University of Illinois at Urbana-Champaign, the Institut de Ci{\`e}ncies de l'Espai (IEEC/CSIC), the Institut de F{\'i}sica d'Altes Energies, Lawrence Berkeley National Laboratory, the Ludwig-Maximilians Universit{\"a}t M{\"u}nchen and the associated Excellence Cluster Universe, the University of Michigan, the National Optical Astronomy Observatory, the University of Nottingham, The Ohio State University, the University of Pennsylvania, the University of Portsmouth, SLAC National Accelerator Laboratory, Stanford University, the University of Sussex, Texas A\&M University, and the OzDES Membership Consortium.

Based in part on observations at Cerro Tololo Inter-American Observatory, National Optical Astronomy Observatory, which is operated by the Association of  Universities for Research in Astronomy (AURA) under a cooperative agreement with the National Science Foundation.

The DES data management system is supported by the National Science Foundation under Grant Numbers AST-1138766 and AST-1536171. The DES participants from Spanish institutions are partially supported by MINECO under grants AYA2015-71825, ESP2015-66861, FPA2015-68048, SEV-2016-0588, SEV-2016-0597, and MDM-2015-0509, some of which include ERDF funds from the European Union. IFAE is partially funded by the CERCA program of the Generalitat de Catalunya. Research leading to these results has received funding from the European Research Council under the European Union's Seventh Framework Program (FP7/2007-2013) including ERC grant agreements 240672, 291329, and 306478. We  acknowledge support from the Australian Research Council Centre of Excellence for All-sky Astrophysics (CAASTRO), through project number CE110001020, and the Brazilian Instituto Nacional de Ci\^enciae Tecnologia (INCT) e-Universe (CNPq grant 465376/2014-2).

This manuscript has been authored by Fermi Research Alliance, LLC under Contract No. DE-AC02-07CH11359 with the U.S. Department of Energy, Office of Science, Office of High Energy Physics. The United States Government retains and the publisher, by accepting the article for publication, acknowledges that the United States Government retains a non-exclusive, paid-up, irrevocable, world-wide license to publish or reproduce the published form of this manuscript, or allow others to do so, for United States Government purposes.

We are grateful for the extraordinary contributions of our CTIO colleagues and the DECam Construction, Commissioning and Science Verification teams in achieving the excellent instrument and telescope conditions that have made this work possible. The success of this project also relies critically on the expertise and dedication of the DES Data Management group.

Funding for the Sloan Digital Sky Survey IV has been provided by the Alfred P. Sloan Foundation, the U.S. Department of Energy Office of Science, and the Participating Institutions. SDSS-IV acknowledges support and resources from the Center for High-Performance Computing at the University of Utah. The SDSS web site is www.sdss.org.

SDSS-IV is managed by the Astrophysical Research Consortium for the Participating Institutions of the SDSS Collaboration including the Brazilian Participation Group, the Carnegie Institution for Science, Carnegie Mellon University, the Chilean Participation Group, the French Participation Group, Harvard-Smithsonian Center for Astrophysics, Instituto de Astrof\'isica de Canarias, The Johns Hopkins University, Kavli Institute for the Physics and Mathematics of the Universe (IPMU) / University of Tokyo, Lawrence Berkeley National Laboratory, Leibniz Institut f\"ur Astrophysik Potsdam (AIP),  Max-Planck-Institut f\"ur Astronomie (MPIA Heidelberg), Max-Planck-Institut f\"ur Astrophysik (MPA Garching), Max-Planck-Institut f\"ur Extraterrestrische Physik (MPE), National Astronomical Observatories of China, New Mexico State University, New York University, University of Notre Dame, Observat\'ario Nacional / MCTI, The Ohio State University, Pennsylvania State University, Shanghai Astronomical Observatory, United Kingdom Participation Group,Universidad Nacional Aut\'onoma de M\'exico, University of Arizona, University of Colorado Boulder, University of Oxford, University of Portsmouth, University of Utah, University of Virginia, University of Washington, University of Wisconsin, Vanderbilt University, and Yale University.

Facilities: DES, LCOGT, Sloan

%%%%%%%%%%%%%%%%%%%%%%%%%%%%%%%%%%%%%%%%%%%%%%%%%%

%%%%%%%%%%%%%%%%%%%%%%%%%%%%%%%%%%%%%%%%%%%%%%%%%%

\section*{Data availability}

The data underlying this article will be shared on reasonable request to the corresponding author. The light curves of J0252 are available in the article's online supplementary material.



\bibliographystyle{mnras}
%\bibliography{binaryrefs}
\input{bbh_v9.bbl}

%\subsection{Code Availability} 

%The code used to perform the spectral fitting analysis of the eBOSS spectrum \textsf{qsofit}\citep{Shen2019} is publicly available at https://github.com/legolason/PyQSOFit.

%\subsection{Data Availability} 

%The optical photometry from the SDSS Stripe 82 and PTF surveys are publicly available at http://sdss.org and http://www.ptf.caltech.edu/. The eBOSS spectrum is publicly available at http://www.sdss.org. The optical light curves from the DES project and the best-fit spectral model to the eBOSS spectrum are available upon reasonable request.

\appendix

\section{Details on archival photometries}\label{sec:archive}

The publicly available PTF photometry was in PTF $g$ and $R$ bands in Vega mags. For consistency we have converted them to the SDSS g and r bands in AB mags following the empirically calibrated relations based on PTF stars (Equations 4 and 5 of \citealt{Ofek2012}). We have further applied the SDSS-DES corrections listed in Equation \ref{eq:kcorrection} for the PTF photometry to be on the DES system. The PTF-to-SDSS correction depends on the (r-i) and (g-r) colors which are variable, however, on the timescales of a few yr. We have adopted the median colors averaged in the last year of the SDSS light curves and the first year of the DES observations that bracketed the PTF $R$-band observations. Furthermore, the current version of the PTF photometric pipeline uses MAG\_AUTO (not aperture or PSF magnitudes) which adjusts the aperture used to extract the source magnitude for each object. This introduces biases in the magnitudes for sources near the survey detection limit such as \obj . The resulting effect on the color correction is a systematic bias toward larger negative values of $r_{{\rm SDSS}}$-R$_{{\rm PTF/SDSS}}$ starting around $r_{{\rm SDSS}}$ of magnitude 19.5 (Figure 2 of \citealt{Ofek2012}). We have empirically corrected for this systematic bias using the median value inferred for sources with similar luminosities of \obj\ (i.e., at $r_{{\rm SDSS}}{\sim}$21 mag). We have further verified the empirical correction by comparing the four PTF $R$-band data points that overlapped with the DES Y1 observations (i.e., around MJD of 56,600), finding a general consistency. Nevertheless, given these significant uncertainties and caveats in the magnitude conversion, as well as the fact that \obj\ is already at the PTF survey detection limit, we do not include the PTF photometry in our baseline analysis. 

The PS1 griz filters are similar to those of the SDSS. We apply the PS1-to-SDSS correction using a third order polynomial provided by \citet{Finkbeiner2016} that shifts the photometry to the SDSS system. The correction depends on the (g-i) color. The color is determined by averaging over PS1 light curve. After the correction to the SDSS system, Eq (1) is then applied for the conversion between the SDSS and DES systems. For ZTF, the photometry has been calibrated to the PS1 system \citep{Masci2019}. We thus follow the same steps in the PS correction and correct the light curves to be on the DES system. The LCOGT filters are similar to the SDSS. We convolve each quasar spectrum with the DES and LCOGT filter transmission curves to calculate the synthetic magnitude difference and correct the LCOGT to be on the DES system.

%%%%%%%%%%%%%%%%%%%%%%%%%%%%%%%%%%%%%%%%%%%%%%%%%%

\section*{affliations}  
$^{1}$Department of Astronomy, University of Illinois at Urbana-Champaign, Urbana, IL 61801, USA \\
 %2
$^{2}$National Center for Supercomputing Applications, University of Illinois at Urbana-Champaign, 605 East Springfield Avenue, Champaign, IL 61820, USA \\
 %3
$^{3}$School of Mathematics and Physics, University of Queensland, QLD 4072 Australia \\
 %4
$^{4}$Kavli Institute for Cosmological Physics, University of Chicago, Chicago, IL 60637, USA \\
 %5
$^{5}$Department of Astronomy and Astrophysics, University of Chicago, 5640 South Ellis Avenue, Chicago, IL 60637, USA \\
 %6
$^{6}$Department of Astronomy, The Ohio State University, 140 West 18th Avenue, Columbus, OH 43210, USA \\
 %7
$^{7}$Center for Cosmology and Astro-Particle Physics, The Ohio State University, 191 West Woodfuff Avenue, Columbus OH, 43210, USA \\
 %8
$^{8}$Institute of Astronomy, University of Cambridge, Madingley Road, Cambridge CB3 0HA, UK \\
 %9
$^{9}$Kavli Institute for Cosmology, University of Cambridge, Madingley Road, Cambridge CB3 0HA, UK \\
 %10
$^{10}$Fermi National Accelerator Laboratory, P. O. Box 500, Batavia, IL 60510, USA \\
 %11
$^{11}$Institute of Cosmology and Gravitation, University of Portsmouth, Portsmouth, PO1 3FX, UK \\
 %12
$^{12}$LSST, 933 North Cherry Avenue, Tucson, AZ 85721, USA \\
 %13
$^{13}$CNRS, UMR 7095, Institut d'Astrophysique de Paris, F-75014, Paris, France \\
 %14
$^{14}$Sorbonne Universit\'es, UPMC Univ Paris 06, UMR 7095, Institut d'Astrophysique de Paris, F-75014, Paris, France \\
 %15
 $^{15}$ Department of Physics \& Astronomy, University College London, Gower Street, London, WC1E 6BT, UK \\
 %16
 $^{16}$ Centro de Investigaciones Energ\'eticas, Medioambientales y Tecnol\'ogicas (CIEMAT), Madrid, Spain \\
 %17
 $^{17}$ Laborat\'orio Interinstitucional de e-Astronomia - LIneA, Rua Gal. Jos\'e Cristino 77, Rio de Janeiro, RJ - 20921-400, Brazil \\
 %18
 $^{18}$ Institut de F\'{\i}sica d'Altes Energies (IFAE), The Barcelona Institute of Science and Technology, Campus UAB, 08193 Bellaterra (Barcelona) Spain \\
 %19
 $^{19}$ Institut d'Estudis Espacials de Catalunya (IEEC), 08034 Barcelona, Spain \\
 %20
 $^{20}$ Institute of Space Sciences (ICE, CSIC),  Campus UAB, Carrer de Can Magrans, s/n,  08193 Barcelona, Spain \\
 %21
 $^{21}$ Kavli Institute for Particle Astrophysics \& Cosmology, P. O. Box 2450, Stanford University, Stanford, CA 94305, USA \\
 %22
 $^{22}$ Department of Physics and Astronomy, University of Pennsylvania, Philadelphia, PA 19104, USA \\
 %23
 $^{23}$ Observat\'orio Nacional, Rua Gal. Jos\'e Cristino 77, Rio de Janeiro, RJ - 20921-400, Brazil \\
 %24
 $^{24}$ Department of Physics, IIT Hyderabad, Kandi, Telangana 502285, India \\
 %25
 $^{25}$ Department of Astronomy/Steward Observatory, 933 North Cherry Avenue, Tucson, AZ 85721-0065, USA \\
 %26
 $^{26}$ Jet Propulsion Laboratory, California Institute of Technology, 4800 Oak Grove Dr., Pasadena, CA 91109, USA \\
 %27
 $^{27}$ Department of Astronomy, University of Michigan, Ann Arbor, MI 48109, USA \\
 %28
 $^{28}$ Department of Physics, University of Michigan, Ann Arbor, MI 48109, USA \\
 %29
 $^{29}$ Instituto de Fisica Teorica UAM/CSIC, Universidad Autonoma de Madrid, 28049 Madrid, Spain \\
 %30
 $^{30}$ Centre for Astrophysics \& Supercomputing, Swinburne University of Technology, Victoria 3122, Australia \\
 %31
 $^{31}$ SLAC National Accelerator Laboratory, Menlo Park, CA 94025, USA \\
 %32
 $^{32}$ Department of Physics, ETH Zurich, Wolfgang-Pauli-Strasse 16, CH-8093 Zurich, Switzerland \\
 %33
 $^{33}$ Santa Cruz Institute for Particle Physics, Santa Cruz, CA 95064, USA \\
 %34
 $^{34}$ Department of Physics, The Ohio State University, Columbus, OH 43210, USA \\
 %35
 $^{35}$ Max Planck Institute for Extraterrestrial Physics, Giessenbachstrasse, 85748 Garching, Germany \\
 %36
 $^{36}$ Universit\"ats-Sternwarte, Fakult\"at f\"ur Physik, Ludwig-Maximilians Universit\"at M\"unchen, Scheinerstr. 1, 81679 M\"unchen, Germany \\
 %37
 $^{37}$ Harvard-Smithsonian Center for Astrophysics, Cambridge, MA 02138, USA \\
 %38
 $^{38}$ Australian Astronomical Optics, Macquarie University, North Ryde, NSW 2113, Australia \\
 %39
 $^{39}$ Departamento de F\'isica Matem\'atica, Instituto de F\'isica, Universidade de S\~ao Paulo, CP 66318, S\~ao Paulo, SP, 05314-970, Brazil \\
 %40
 $^{40}$ George P. and Cynthia Woods Mitchell Institute for Fundamental Physics and Astronomy, and Department of Physics and Astronomy, Texas A\&M University, College Station, TX 77843,  USA \\
 %41
 $^{41}$ Instituci\'o Catalana de Recerca i Estudis Avan\c{c}ats, E-08010 Barcelona, Spain \\
 %42
 $^{42}$ School of Physics and Astronomy, University of Southampton,  Southampton, SO17 1BJ, UK \\
 %43
 $^{43}$ Cerro Tololo Inter-American Observatory, National Optical Astronomy Observatory, Casilla 603, La Serena, Chile \\
 %44
 $^{44}$ Brandeis University, Physics Department, 415 South Street, Waltham MA 02453 \\
 %45
 $^{45}$ Instituto de F\'isica Gleb Wataghin, Universidade Estadual de Campinas, 13083-859, Campinas, SP, Brazil \\
 %46
 $^{46}$ Computer Science and Mathematics Division, Oak Ridge National Laboratory, Oak Ridge, TN 37831 \\
 %47
 $^{47}$ Argonne National Laboratory, 9700 South Cass Avenue, Lemont, IL 60439, USA \\

% Don't change these lines
\bsp	% typesetting comment
\label{lastpage}
\end{document}